\documentclass[twocolumn]{aastex63}

\newcommand{\msun}{M$_\odot$}
\newcommand{\ms}{m~s$^{-1}$}
\newcommand{\ca}{Ca {\sc ii} H\&K}
\newcommand{\logrhk}{$\log R'_{\rm HK}$}
\newcommand{\eg}{e.g.}
\newcommand{\ff}{\emph{FF'}}
\newcommand{\bhat}{$|\hat{B}_{\rm obs}|$}
\newcommand{\ie}{i.e.}

\newcommand{\kepler}{\emph{Kepler}}
\newcommand{\drv}{$\Delta RV$}

\newcommand{\mearth}{M$_\oplus$}
\newcommand{\rms}{{\sc rms}}
\newcommand{\nrvsdo}{2811}
\newcommand{\nrvsorce}{2535}

\newcommand{\Ppl}{300.38}

\received{\today}
\revised{}
\accepted{}
\submitjournal{ApJ}

\shorttitle{Unsigned magnetic flux and RV variations of the Sun}
\shortauthors{Haywood R. D. et al.}

\begin{document}

\title{Unsigned magnetic flux as a proxy for radial-velocity variations \\ in Sun-like stars}

\correspondingauthor{R. D. Haywood}
\email{rhaywood@cfa.harvard.edu}

\author[0000-0001-9140-3574]{R. D. Haywood}
\affiliation{Center for Astrophysics ${\rm \mid}$ Harvard {\rm \&} Smithsonian, 60 Garden Street, Cambridge, MA 02138, USA}
\affiliation{Astrophysics Group, University of Exeter, Exeter EX4 2QL, UK}
\affiliation{NASA Sagan Fellow}

\author{T. W. Milbourne}
\affiliation{Center for Astrophysics ${\rm \mid}$ Harvard {\rm \&} Smithsonian, 60 Garden Street, Cambridge, MA 02138, USA}
\affiliation{Department of Physics, Harvard University, 17 Oxford Street, Cambridge MA 02138, USA}

\author[0000-0001-7032-8480]{S. H. Saar}
\affiliation{Center for Astrophysics ${\rm \mid}$ Harvard {\rm \&} Smithsonian, 60 Garden Street, Cambridge, MA 02138, USA}

\author[0000-0001-7254-4363]{A. Mortier}
\affiliation{KICC \& Cavendish Laboratory, University of Cambridge, J.J. Thomson Avenue, Cambridge CB3 0HE, UK}

\author[0000-0001-5132-1339]{D. Phillips}
\affiliation{Center for Astrophysics ${\rm \mid}$ Harvard {\rm \&} Smithsonian, 60 Garden Street, Cambridge, MA 02138, USA}

\author[0000-0002-9003-484X]{D. Charbonneau}
\affiliation{Center for Astrophysics ${\rm \mid}$ Harvard {\rm \&} Smithsonian, 60 Garden Street, Cambridge, MA 02138, USA}

\author[0000-0002-8863-7828]{A. Collier Cameron}
\affiliation{SUPA, Institute for Astronomy, Royal Observatory, University of Edinburgh, Blackford Hill, Edinburgh EH93HJ, UK}

\author[0000-0001-8934-7315]{H. M. Cegla}
\affiliation{University of Warwick, Department of Physics, Gibbet Hill Road, Coventry, CV4 7AL, UK}
\affiliation{Observatoire de Gen\`{e}ve, Universit\'{e} de Gen\`{e}ve, 51 Chemin des Maillettes, CH-1290 Versoix, Switzerland}

\author{N. Meunier}
\affiliation{Universit\'{e} Grenoble Alpes, CNRS, IPAG, F-38000 Grenoble, France}

\author[0000-0002-4677-8796]{M. L. Palumbo III}
\affiliation{Pennsylvania State University, State College, PA 16801, USA}

\begin{abstract}

We estimate disc-averaged RV variations of the Sun over the last magnetic cycle, from the single Fe {\sc I} line observed by SDO/HMI, using a physical model for rotationally modulated magnetic activity that was previously validated against HARPS-N solar observations. We estimate the disc-averaged, unsigned magnetic flux and show that a simple linear fit to it reduces the \rms~of RV variations by 62\%, \ie~a factor of 2.6. 
We additionally apply the \ff~method, which predicts RV variations based on a star's photometric variations.
At cycle maximum, we find that additional physical processes must be at play beyond suppression of convective blueshift and velocity imablances resulting from brightness inhomogeneities, in agreement with recent studies of solar RV variations. 
By modelling RV variations over the magnetic cycle using a linear fit to the unsigned magnetic flux, we recover injected planets at an orbital period of $\approx$~300~days with RV semi-amplitudes down to 0.3~\ms. 
To reach semi-amplitudes of 0.1~\ms, we will need to identify and model additional physical phenomena that are not well traced by \bhat~or \ff.
The unsigned magnetic flux is an excellent proxy for rotationally modulated, activity-induced RV variations, and could become a key tool in confirming and characterising Earth analogs orbiting Sun-like stars. The present study motivates ongoing and future efforts to develop observation and analysis techniques to measure the unsigned magnetic flux at high precision in slowly rotating, relatively inactive stars like the Sun.

\end{abstract}

\keywords{Sun: activity --- Sun: faculae, plages --- (Sun:) sunspots
--- planets and satellites: detection --- methods: data analysis --- techniques: radial velocities}

\section{Introduction} \label{sec:intro}

The main obstacle we face in detecting, confirming and characterising Neptune- to Earth- mass exoplanets via radial-velocity (RV) monitoring is the intrinsic variability of the host stars themselves (see \citet{nas2018,fischer2016} and references therein). 
RV monitoring is the most widely applicable technique to determine the masses of the small planets to be discovered by TESS and PLATO. 
Mass is the most fundamental parameter of a planet: it is central to theoretical models of planet composition and structure \citep[\eg][]{zeng2013}. Planetary mass dictates the amount of observing time required to characterise a planet's atmosphere, so it is essential that we know masses reliably to plan atmospheric  follow-up observations \citep{morley2017,batalha2019}, \eg~with JWST and ARIEL.
To determine accurate and precise planetary masses, we need to develop robust, physically motivated models for stellar variability.
We still lack a complete and detailed understanding of how the interplay between magnetic fields and granulation gives rise to RV variations on the Sun and other stars \citep{eprvwg}.

The Sun is the only star whose surface we can image directly and at high resolution, making it an ideal test bench to examine the physical phenomena responsible for instrinsic RV variability. It is also the only star whose RV we know independently of spectroscopic measurements (\eg~from HARPS-N).

On timescales of several rotation periods (weeks--months), RV variability is driven by magnetic activity in the photosphere. 
The manifestations of magnetic activity relevant to the present analysis are sunspots and faculae. 
Sunspots are relatively large, dark areas of strong magnetic fields \citep[\eg][Chap. 8]{foukal2004}.
Faculae are small, bright magnetic flux tubes \citep{spruit1976}. They tend to be located in the lanes between supergranular cells and are spread all over the solar surface, thus forming the photospheric magnetic \emph{network} \citep[\eg][Chap. 5, p. 145; Chap. 8]{foukal2004}. In regions of enhanced magnetic activity, faculae cluster into areas of \emph{plage} \citep[\eg][Chap. 1, Fig. 1.1]{schrijver2000}.
The total surface area covered by network varies throughout the Sun's 11-year cycle, although to a much smaller extent than spots and plage \citep{meunier2018,meunier2003}. 
At low activity levels, magnetic elements tend to be spread throughout the solar surface in the form of network.
Plage coverage increases with magnetic activity, and at high activity levels, the majority of magnetic elements on the surface are concentrated in plage, rather than in network (\eg~see plage and network filling factors shown in Figure~\ref{fig_all}). Plage generally decays into network over over timescales of several rotations, which explains the larger network coverage when activity is high.  

Magnetic elements inhibit convective motions, thereby suppressing some of the convective blueshift that results from granulation \citep[\eg][]{Dravins:1981tl}. 
This suppression of convective blueshift is the dominant contributor to RV variations in the Sun \citep{saar1997}. 
Solar observations show RV variations, modulated by the Sun's rotation and evolving over timescales of days to weeks, with amplitudes of several \ms~\citep{meunier2010a,meunier2010b,dumusque2014,haywood2016,milbourne}.
Convective blueshift is suppressed by faculae in concentrated areas of plage, while faculae in the network, being more spatially diffuse, do not perturb convective flows significantly \citep[][Sect. 4.4]{milbourne}. 
Sunspots contribute little to observed suppression of convective blueshift, as they are dark (and thus contribute relatively less to observed spectra) and cover very little area in comparison to faculae  \citep[\eg][]{Lagrange2010I,haywood2016}.
At high solar activity levels, sunspots do contribute significantly to RV variations. Because they are much darker than the surrounding photosphere, they produce significant inhomogeneities in surface brightness, which result in RV variations with an~\rms~ of 60 c\ms~and frequent peak-to-peak amplitudes of 2~\ms, with variations of up to 5~\ms~\citep{Lagrange2010I}. 
On the other hand, faculae (both in network and plage) are only 10\% brighter than the quiet photosphere at optical wavelengths; moreover, they are distributed much more uniformly longitudinally on the solar disk, so their brightness-induced RV contribution mostly cancels out on rotational timescales. On longer (magnetic cycle) timescales, large-scale changes in the number of faculae can produce long term, bulk RV shifts \citep[\eg][Fig.8]{SaarFisch00,meunier2010b}.

Magnetic elements enhance chromospheric column density, which strengthens the emission reversals in the \ca~cores. Thus, the filling factor of magnetic elements correlates strongly with the amount of emission in the cores of the \ca~lines \citep[\eg][Fig.1]{meunier2018} as measured by the \logrhk~index \citep{vaughan1978,noyes1984}. 
Solar observations show that RV variations and \logrhk~correlate strongly over long timescales of several years, \ie~over the Sun's 11-year magnetic cycle \citep[\eg][Fig.13]{meunier2010a}.
However, when we look at shorter timescales of a few weeks to months, \ie~on the solar rotation timescale, the \logrhk~does not systematically trace RV variations down to sub \ms~precision. 
This is the case during both low and high activity phases. 
At low activity levels (on the rotation timescale), the low correlation between \logrhk~and RV variations can be explained by the fact that magnetic elements are predominantly found in network, which contributes to \ca~emission but does not affect RV variations \citep{milbourne}.
At high activity levels (on the rotation timescale), RV variations do not correlate well with \ca~emission, \eg~in \citet[][Fig.12]{Lagrange2010I} and \citet[][Fig.10]{haywood2016}. Part of this discrepancy may be that the \ca~emission forms in the chromosphere, and therefore follows a different limb-darkening law and projection effects than RVs, which are measured from photospheric absorption lines (see Section~\ref{app_hysteresis} in this paper). 
To summarise, solar RV variations induced by magnetic activity on timescales of the order of a few rotation periods are not directly correlated and in phase with \ca~emission, in part because faculae affect RV variations differently depending on whether they are in sparse network or in concentrated regions of plage, and due to changes in spectral line profiles.

We see a similar behaviour in observations of Sun-like stars.
As part of the Mt. Wilson {\sc HK} Project, several dozen slowly rotating, old Sun-like stars were monitored in optical photometry and in Ca {\sc II} emission over the past several decades \citep{Baliunas:1995cc,Wilson:1978is}. Observations showed that as these stars became more magnetically active (as indicated via their S-index), they also got brighter. Their surfaces are therefore dominated by bright faculae rather than dark spots, just like the Sun \citep{Lockwood2007,radick2018}.
On timescales of several years, the \rms~of stellar RV variations increases as the \logrhk increases \citep{SaarFisch00,Lovis:2011vq}.
\citet{Aigrain:2012} developed a model to estimate stellar activity-induced RV variations based on the star's optical photometric variations. The model accounts for RV variations produced by dark spots and bright plage \emph{that are spatially associated with spots}, both through rotational flux imbalance and suppression of convective blueshift. \citet{Haywood2014} tested this model on the solar analog CoRoT-7 using simultaneous photometric and RV observations taken at high cadence over a rotation period. 
While the model of \citet{Aigrain:2012} captures a significant part of the rotationally modulated RV signal, it leaves out an equally significant rotationally modulated signal. 
This additional RV variation likely originates from magnetic regions that have low intensity contrast, and whose brightness-induced RV variations is therefore low.
These observations and their interpretation are consistent with the Sun's behaviour (the solar surface is dominated by low-contrast plage). 

To confirm and characterise long-period, low-mass exoplanets, we need a proxy that traces RV variations systematically and at sub-\ms~precision. 
On both the Sun and other Sun-like stars, \ca~emission does not systematically correlate as strongly with activity-induced RV variations.
For the Sun, a strong correlation has been observed between activity-induced RV variations and the unsigned, full-disc magnetic flux, over the magnetic cycle \citep{deming1994,lanza2016,meunier2018} and on the rotation timescale \citep{haywood2016}. 
The disc-averaged RV timeseries estimated from the Michelson Doppler Imager onboard the Solar and Heliospheric Observatory (SoHo/MDI) by \citet{meunier2010b} and subsequent papers by Meunier et al. is well-sampled and spans over 4 years of Cycle 23, prior to SDO's launch; however, it cannot be compared to ``ground-truth", direct disc-integrated RV observations. To date, systematic RV campaigns of the Sun as a star have been carried out spectroscopically, with HARPS via sunlight reflected from asteroids \citep{haywood2016,lanza2016} and more recently with dedicated solar feeds at HARPS-N \citep{Dumusque:2015,colliercameron} and HARPS \citep[][HELIOS]{dumusque2019}.

In the present analysis, we estimate rotationally modulated RV variations of the Sun from SDO/HMI images over 8 years, at daily cadence (Section~\ref{sec_sdodata}), using a technique that has been validated against direct Sun-as-a-star HARPS-N observations (Section~\ref{sec_reconstruct}). 
We present timeseries of RV variations and unsigned magnetic flux in Section~\ref{sec_reconstruct}.  
We model our RV timeseries using a linear fit in unsigned magnetic flux in Section~\ref{sec_model_b} and with the \ff~model of \citet[][]{Aigrain:2012} in Section~\ref{sec_model_bff}.
In Section~\ref{sec_others}, we identify potential additional physical effects giving rise to rotationally modulated RV variations that are not well traced by either models, and identify limitations of the \ff~model.
We perform simple planet injections to assess the performance of the unsigned magnetic flux for mitigating rotationally modulated RV variations in Section~\ref{sec_planet}. 
We discuss future prospects, including ways to measure the unsigned magnetic flux in other stars in Section~\ref{sec_prospects}, and present our conclusions in Section~\ref{sec_conclusion}.

\section{data} \label{sec_data}
\subsection{SDO/HMI images}\label{sec_sdodata}

\begin{figure*}[t]
\centering
\includegraphics[scale=0.56]{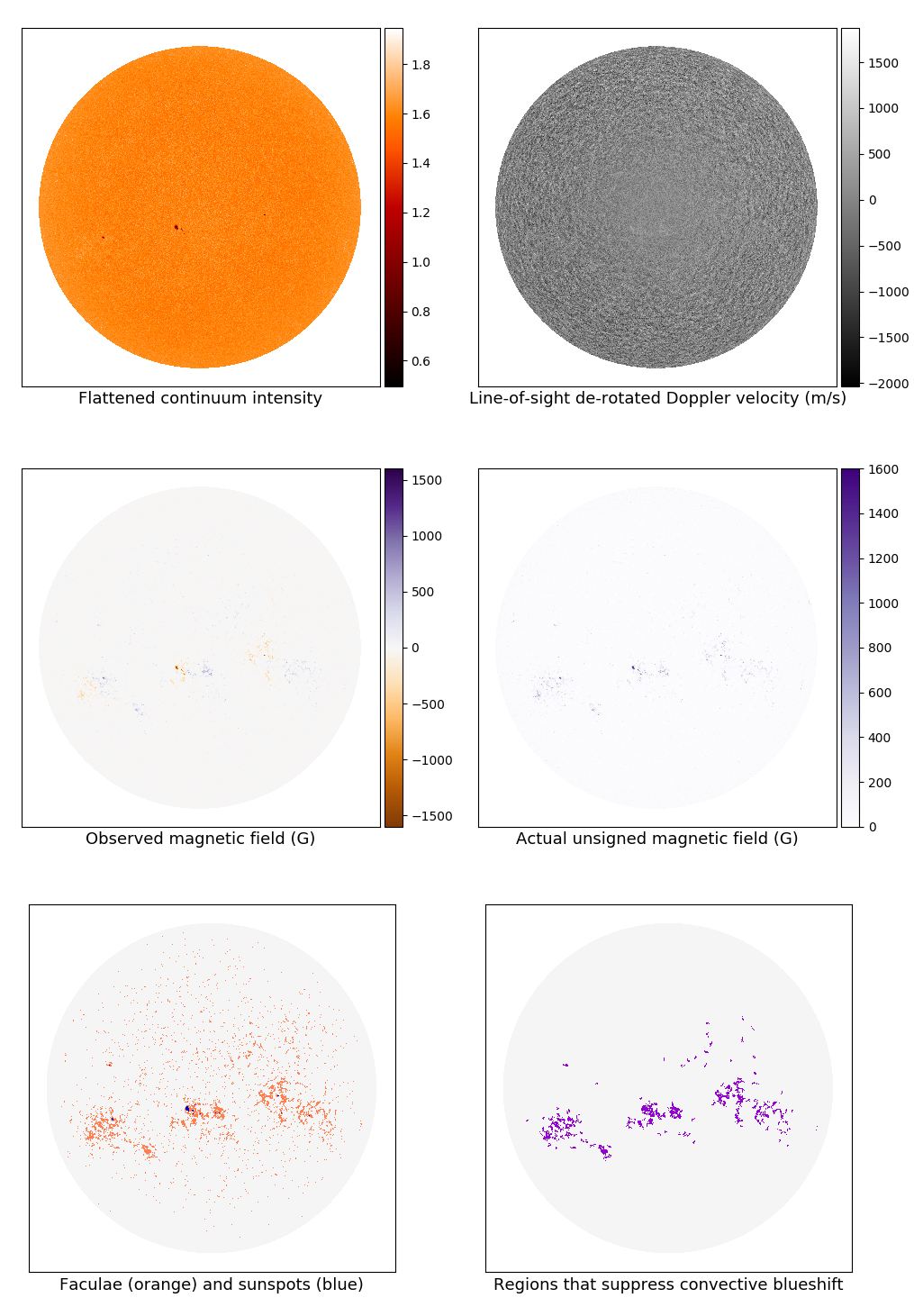}
\caption{SDO/HMI data products for an observation set taken on 2015 November 28 at 20:00:00 UTC. This set of images is representative of the Sun during high activity levels. Faculae and sunspots fill 3.25\% and 0.03\% of the solar disc, respectively. The areas that suppress convective blueshift, predominantly faculae in regions of plage, shown in the last panel, fill 1.59\% of the solar disc. The disc-averaged unsigned magnetic flux is 9.99~G.
\emph{Notes:} The flattened intensity is normalised to the mean intensity. The line-of sight velocity is shown after subtracting the solar rotation profile and the velocity of the spacecraft.  \label{fig_pictures}}
\end{figure*}

We use 720-second HMI exposures of continuum intensity (both uncorrected and corrected for limb darkening by the HMI team), Dopplergrams, and magnetograms, as represented in Figure~\ref{fig_pictures}. 
The HMI instrument takes 6 measurements of intensity across a narrow wavelength range centered on the Fe~{\sc I} line at 6173~\AA~\citep[see Fig. 6 of][]{schou2012}. 
These points are fitted with a Gaussian profile to generate the main HMI data products, to generate the main HMI data products including velocity (line shift), intensity (depth), magnetic field strength (width due to Zeeman broadening) and continuum intensity \citep[][Sect.3.3]{schou2012}. 
The line shifts and magnetic field values extracted for each pixel should be independent, physical quantities, obtained from combinations of the different intensities at different points on the measured Fe~{\sc I} line. 
While the line asymmetry stemming from convection within each pixel is not preserved, we expect this technique to capture asymmetries due to physical processes occuring over scales larger than a pixel.
We refer the reader to \cite{schou2012} for further details on how these images are extracted from the raw filtergrams. 

\subsubsection{Temporal sampling}
SDO/HMI has operated almost continuously since the start of the mission, except for spacecraft operations and calibrations, and eclipses that happen due to the geosynchronous orbit of the SDO spacecraft \citep{hoeksema}. 
There have been very few anomalies requiring interruptions, none of which have been prolonged compared to the seasonal eclipses. 
We take a set of images every 4 hours (6 times per 24-hour period) from 2010 April 07, 04:00:00 UTC up to 2018 January 12 20:00:00 UTC, amounting to a total of 16855 sets of images spanning \nrvsdo~days.
We take daily averages to minimise the contribution of short-term processes, namely oscillations and magnetoconvection.
We choose not to use the SDO/HMI images at their highest cadence in order to maintain the relevance of this analysis to stellar studies, while still sampling the Sun multiple times a day. Indeed, the cadence of current and planned stellar RV surveys is 1-3 observations per night at most.

\subsubsection{Instrument precision and stability}\label{sec_dopplergrams}

\begin{table*}[t]
    \centering
    \begin{tabular}{p{6cm}p{2cm}p{8.0cm}}
        Quantity & Measured \hskip3ex to precision & Notes \& References \\
        
        \rule{0pt}{0ex} \\
        \hline
        \rule{0pt}{0ex} \\
        
        LOS velocity per pixel, $v_{\rm pix}$ & 7 \ms &  Photon noise at disk center in Dopplergram (hmi.V\_720s). From \cite[][Table 1]{couvidat}.  \\ 

        LOS velocity over full solar disc & 0.002 \ms & Average over the full timeseries of:
        $v_{\rm pix}/\sqrt{n_{\rm pix}}$, where $n_{\rm pix}$ is the number of pixels within
        $\mu = 0.3$ in a given Dopplergram. \\
        
        Spacecraft velocity & 0.01 \ms & From \cite{couvidat}. \\
        
        Pipeline to measure $\Delta RV$ from HMI images & 0.1 \ms & Systematic uncertainty in our analysis, particularly in classification of active-region areas. \\
        
        {\bf LOS velocity (disc-averaged)} & {\bf 0.1 \ms} & We add the instrument, pipeline and astrophysical uncertainties listed above in quadrature. \\
    
        \rule{0pt}{0ex} \\
        \hline
        \rule{0pt}{0ex} \\    
        
        LOS unsigned magnetic flux per pixel, $B_{\rm pix}$ & 3 G & Photon noise at disk center in magnetogram (hmi.M\_720s). From \cite[][Table 1]{couvidat}. \\
        
        {\bf LOS unsigned magnetic flux (disc-averaged)} & {\bf 0.0009 G} & Average over the full timeseries of: $B_{\rm pix}/\sqrt{n_{\rm pix}}$. \\
        
        \rule{0pt}{0ex} \\
        
        \hline
    \end{tabular}
    
    \rule{0pt}{0ex} \\

    \caption{Uncertainties for the quantities estimated in this analysis, namely the line-of-sight (LOS) velocity and the unsigned LOS magnetic flux. See details in Section~\ref{sec_dopplergrams}.}

    \label{table_budget}
\end{table*}

\begin{figure}[]
\centering
\includegraphics[scale=0.43]{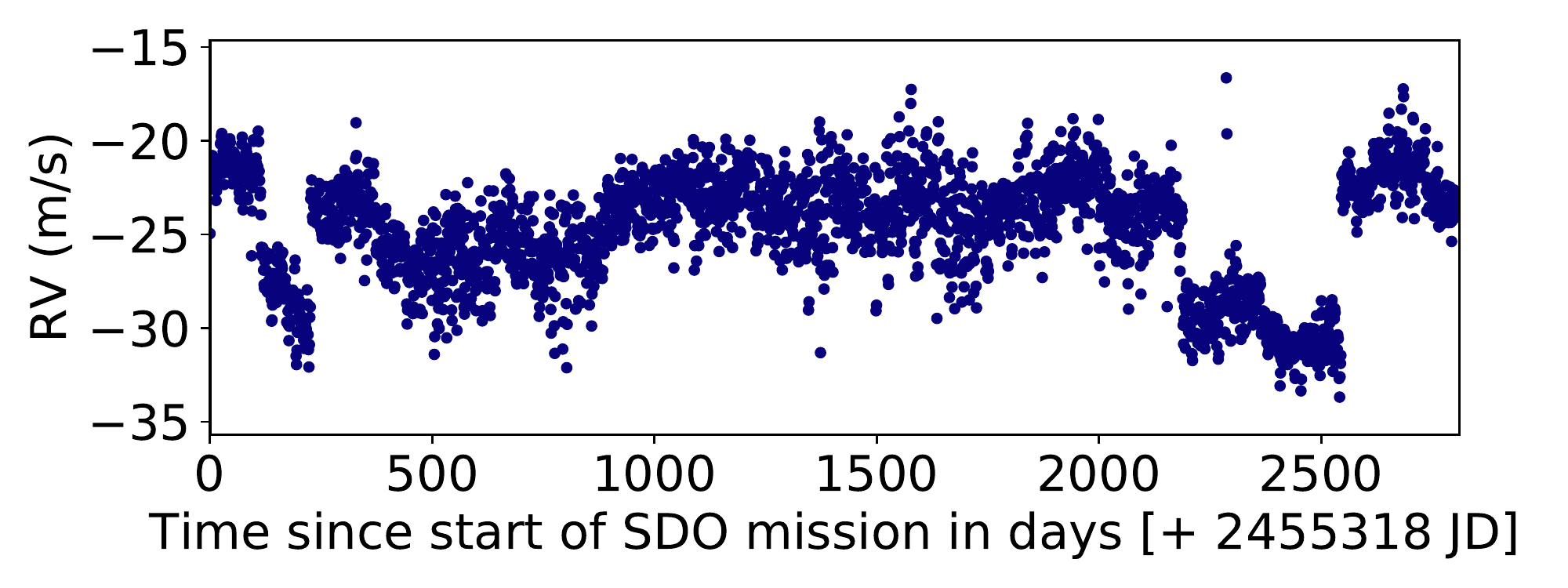}
\caption{Disc-averaged radial velocity of the Sun as estimated from SDO/HMI images, before subtracting the disc-averaged velocity of the quiet Sun. The jumps in RV are instrument systematics, as HMI is not calibrated for long-term stability.  \label{fig_rv_hat}}
\end{figure}

All sources of uncertainty that affect our RV estimates are listed in Table~\ref{table_budget}.
\cite{hoeksema} recently assessed the performance of HMI and reported that the instrument continues to work according to its original specification. The data products are corrected on a regular basis as the calibrations improve (\eg~instrument thermal environment, focus, image distortions, optics alignment, cosmic rays correction, \emph{etc}.). They report that the quality of the data is very uniform with time. 
The HMI data products are well calibrated \citep{hoeksema}, with the exception of the long-term stability of the Dopplergrams. 
The Doppler velocity maps were designed for helioseimology investigations, so they are not calibrated to be stable over timescales longer than a few hours or days \citep{Scherrer2011}. This is shown in Figure~\ref{fig_rv_hat}, which shows the disc-averaged velocity of the Sun taking irregular jumps of several \ms~over the course of the SDO mission. To correct for this effect, we perform all of our velocity calculations relative to the disc-averaged velocity of the magnetically inactive, quiet Sun for each Doppler image \citep[as in][]{meunier2010b,haywood2016,milbourne}. We estimate the velocity of the quiet Sun by summing over all pixels identified as non-magnetic (see Section~\ref{sec_reconstruct}), and excluding pixels considered to magnetically active (even at the peak of the Sun's magnetic cycle, fewer than 5\% of all pixels within the solar disc are magnetically active). 
We are therefore only considering RV \emph{variations} (\drv). 
Importantly, this subtraction cancels out all velocity flows from the quiet Sun. This means that our RV estimates are free from pressure-mode (p-mode) oscillations, granulation and supergranulation motions, which would otherwise induce uncorrelated noise at the 1~\ms~level \citep[\eg][]{meunier2015}.

Beyond the lack of long-term calibration, \cite{couvidat} report that the most significant source of instrument-related uncertainty that remains in individual HMI Dopplergrams is the orbital velocity of the spacecraft, that is uncertain to 0.01~\ms. 
Indeed, we see a systematic sinusoidal shift with a periodicity of 12 and 24 hours due to the spacecraft's orbit. 
This systematic should mostly average out through our sampling, but to be on the safe side, we add \citet{couvidat}'s uncertainty of 0.01~\ms~in quadrature to our RV uncertainties (see Table~\ref{table_budget}).
The total number of pixels inside the solar disc as it appears on the HMI image varies by $\pm 3\%$ over the course of each year, primarily due to Earth's eccentric orbit. To correct for this effect, we normalise all our quantities by the total number of pixels in each image, \ie~we estimate disc-averaged quantities.
We ran tests to assess the potential effect of hypothetical spurious pixels from the continuum, Doppler and magnetic images.
For example, we find that setting the velocity to zero for a large patch of 20 $\times$ 20 pixels square incurs changes in RV of 0.04~\ms. The RV effect of spurious pixels distributed randomly on the solar disc is less than 0.01~\ms.
To be conservative, we assign a constant RV uncertainty of 0.1~\ms~for all RV estimates, to account for uncertainties arising from our  pipeline to estimate RVs (detailed in Section~\ref{sec_reconstruct}). Our choice of magnetic and continuum intensity thresholds are based on previous studies, but they remain somewhat arbitrary, and changing them slightly will impact RV estimates at the c\ms~level. 
We conservatively remove sets of images with $v \sin i$ values more than 3-sigma deviant from the mean (6 out of 16855), and those with focal length values more than 3-sigma deviant (2 out of 16855).

\subsection{SORCE total solar irradiance observations}\label{sec_tsi}
To apply the \ff~method of \citet[][]{Aigrain:2012} in Section~\ref{sec_model_bff}, we use total solar irradiance observations (TSI; akin to a \kepler~lightcurve for the Sun) taken by the Total Irradiance Monitor (TIM) onboard the SOlar Radiation and Climate Experiment (SORCE) satellite \citep{2001AGUSM..SH52A08K,2005SoPh..230...27L}\footnote{SORCE TSI data available at: \url{http://lasp.colorado.edu/home/sorce/data/tsi-data/}}.
We use the TIM's daily average TSI measurements, which span the full duration of our SDO/HMI timeseries. The TIM takes observations every 50 seconds when the spacecraft faces the Sun, and these observations are then combined to produce daily averages. We concatenate this timeseries with our SDO/HMI timeseries of daily images. 
Because there is a gap in the SORCE timeseries around 1200-1400 days into the SDO mission, we are left with a combined timeseries of \nrvsorce~daily observations, spanning \nrvsdo~days. 
The SORCE timeseries is plotted in the last panel of Figure~\ref{fig_all}.
The TIM achieves a precision of 4-17 ppm per observation \citep{kopp2014}. For comparison, \kepler~achieved the same level of precision on a 7 to 9th V magnitude star in a long-cadence observation\footnote{Source: \\ \url{https://keplergo.arc.nasa.gov/CalibrationSN.shtml}} (30-minute integration time). 
TIM has a long-term stability of about 10 ppm per year \citep{kopp2014}.

\subsection{S-index from Mt Wilson and HARPS-N}
We compare our SDO/HMI-derived quantities (\drv, \bhat) against \ca~emission observations. For this, we use overlapping S-index observations of the Sun seen as a star at the Mount Wilson Observatory as part of the HK Project, fully homogenised and calibrated by \cite{egeland2017}. Their observations run from 1966 until 2015.
To cover the 2015-2018 period, we use daily averaged S-index observations taken by the solar telescope that feeds the HARPS-N spectrograph since July 2015 \citep{milbourne}.
The Mt Wilson and HARPS-N datasets overlap for around 45 days in 2015, which we can use to stitch the two S-index timeseries together. We do this by rescaling the overlapping part of the HARPS-N S-index timeseries so it has the same variance as the Mt Wilson S-index timeseries and subtracting the offset between the two datasets. The full timeseries is shown in the second to last panel of Figure~\ref{fig_all}.


\section{Estimating the full-disc RV variations and magnetic flux of the Sun} \label{sec_reconstruct}

We estimate disc-averaged active-region filling factors, RV variations and unsigned (unpolarised) magnetic fluxes of the Sun using spatially resolved images from SDO/HMI images according to the same method as \citet{milbourne}, adapted from that of \citet{haywood2016}, which builds on the techniques originally developed by \citet{meunier2010b} and \citet{fligge2000}.

\subsection{Separating magnetically active regions from quiet Sun}
We separate magnetically active regions from quiet-Sun regions by applying a threshold in unsigned radial magnetic field strength for each pixel according to the cutoff found by \citet{yeo2013}:
\begin{equation}
|B_{{\rm r}, ij}| > 3\, \sigma_{B_{{\rm obs}, ij}} / \mu_{ij},
\end{equation}
\label{eqn_1}
The factor $\mu$ accounts for foreshortening, and is equal to $\cos \theta_{ij}$, where $\theta_{ij}$ is the angle between the outward normal to the feature on the solar surface and the direction of the line-of-sight of the SDO spacecraft.
The term $\sigma_{B_{{\rm obs}, ij}}$ represents the noise in the observed magnetic field, for each pixel at position $i,j$ on the image. \citet{yeo2013} estimated $\sigma_{B_{{\rm obs}, ij}}$ to be 8~G (photon-dominated), so the magnetic field threshold is 24~G. 
We exclude \emph{isolated} pixels that are above this threshold as they are likely to be false positives.

\subsection{Filling factors of sunspots \& plage}
To identify faculae and sunspots, we apply the intensity threshold of \citet{yeo2013}, at 0.89$\,$ times the mean flattened intensity over quiet-Sun regions.
We further identify faculae in concentrated regions of \emph{plage}, as opposed to faculae dispersed in the \emph{network} (\emph{cf.} Introduction, \S~3), according to the area threshold estimated by \citet{milbourne}. We identify plage as magnetically active facular regions whose area on the flattened solar disc exceeds 20 microhemispheres ($\mu$hem), corresponding to about 60~Mm$^2$.

We estimate the disc-averaged filling factors of sunspots and plage as follows:
\begin{equation}
f_{\rm spot, plage} = \frac{1}{N_{\rm pix}} \, \sum_{ij} W_{ij},
\end{equation}\label{eqn_ff}
where $N_{\rm pix}$ is the total number of pixels in the solar disc and the weight $W_{ij}$ is set to 1 in sunspot (or plage) pixels, and 0 in quiet-Sun pixels.

\subsection{RV variations}\label{sec_reconstruct_model}
\cite{milbourne} derived solar RV variations from SDO/HMI images for an 800-day period overlapping disc-integrated RV observations of the Sun with the HARPS-N spectrograph. They  reproduced rotation-modulated RV variations in good agreement with the HARPS-N observations, down to an \rms~level of 1.21~\ms, which is consistent with residual motions that are expected from granulation and supergranulation \citep{meunier2015}.
Their model, which we apply here, accounts for the suppression of convective blueshift from magnetic regions, and the velocity imbalances resulting from brightness inhomogeneities.
We refer the reader to the Appendix of \citet{milbourne}, which fully describes the model. Estimating RV variations according to this technique and model allows us to determine solar RV variations that we can compare directly with spectroscopic measurements of other stars, which are derived from thousands of spectral lines, not just the Fe~{\sc i} line measured by SDO/HMI.

\subsection{Unsigned magnetic flux $|\hat{B}_{\rm obs}|$}
We compute the disc-averaged, line-of-sight unsigned (\ie~unpolarised) magnetic flux of the Sun, by summing the intensity-weighted line-of-sight absolute magnetic flux in each pixel according to \citet{haywood2016}:
\begin{equation}
|\hat{B}_{\rm obs}| = \frac{\sum_{ij} |B_{{\rm obs}, ij}| \, I_{ij}}{\sum_{ij} I_{ij}}
\end{equation}
where $I_{ij}$ is the observed, non-flattened continuuum intensity of the Sun. We do not flatten the intensity continuum in order to obtain the \emph{observed} unsigned magnetic flux.

\subsection{Correlations between \drv, \bhat, photometry and S-index}

\begin{figure*}[t]
\centering
\includegraphics[scale=0.75]{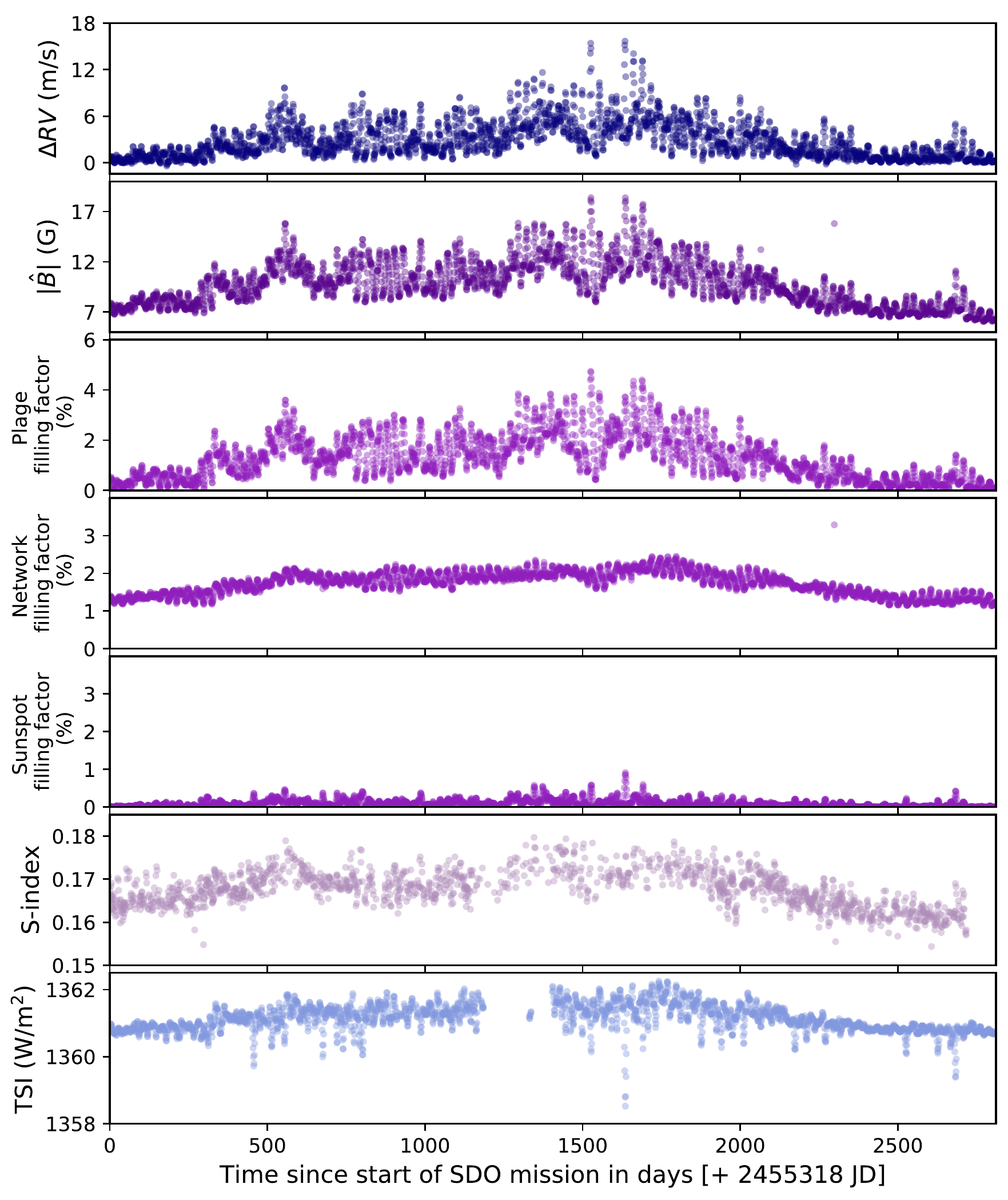}
\caption{Top to bottom: SDO/HMI full-disk quantities: total RV \emph{variations} (after subtracting quiet-Sun velocity), observed unsigned magnetic flux; filling factors of faculae in concentrated areas of plage, faculae in the diffuse network, and sunspots; \ca~emission S-index from Mt Wilson \citep{egeland2017} and HARPS-N \citep{milbourne}; total Solar Irradiance (TSI) from SORCE.  \label{fig_all}}
\end{figure*}

\begin{figure*}[t]
\centering
\includegraphics[scale=0.6]{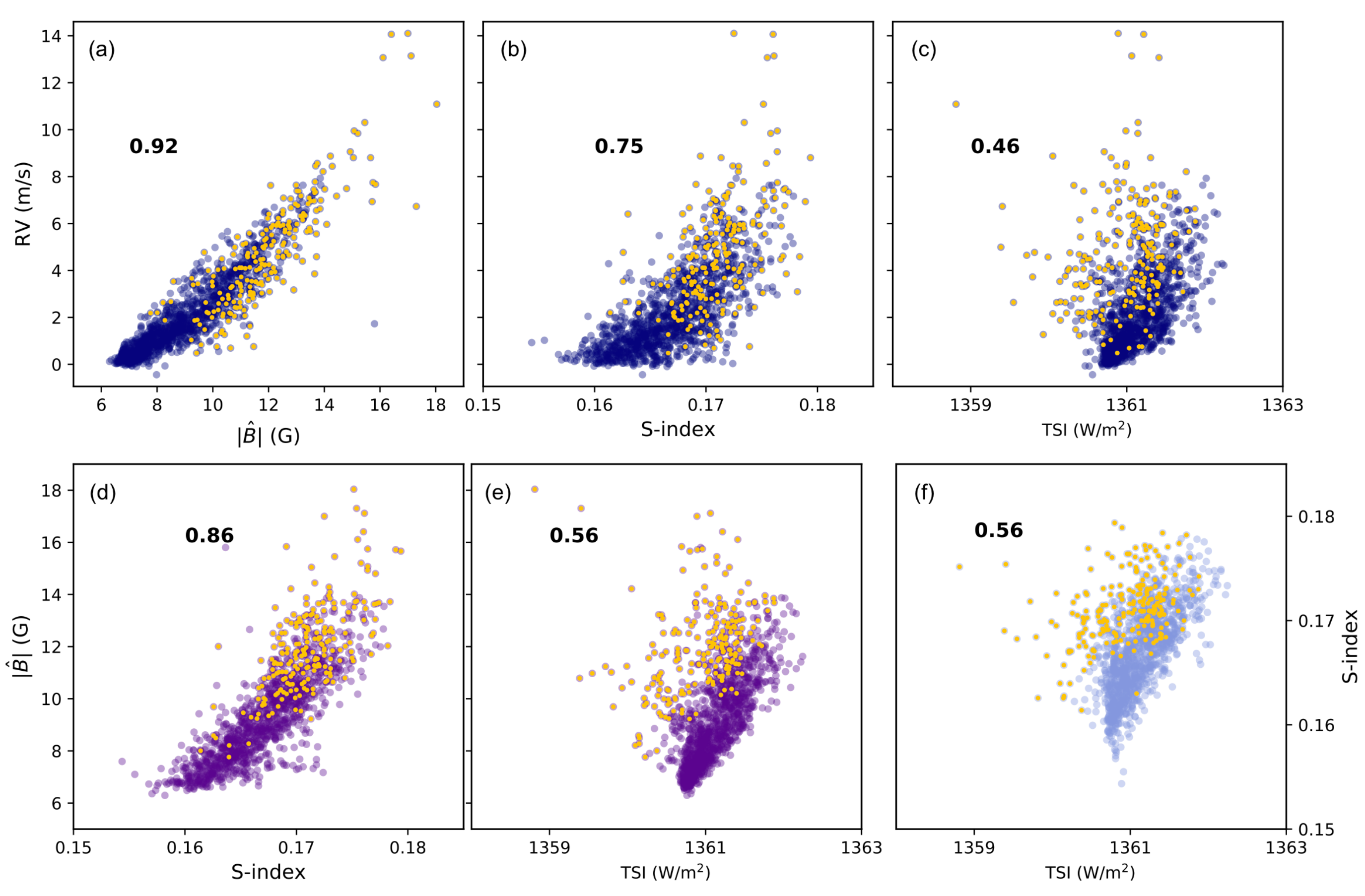}
\caption{Top: SDO/HMI-derived RV variations plotted as a function of (a) unsigned magnetic flux \bhat, (b) S-index and (c) TSI. Bottom: unsigned magnetic flux as a function of (d) S-index and (e) TSI; and (f): S-index and TSI against one another. Observations highlighted in yellow have a ``high" sunspot filling factor ($f_{\rm spot} > 0.15$\%). Spearman correlation coefficients are given for each pair of correlates.\label{fig_correlations}}
\end{figure*}

The timeseries of RV variations, unsigned magnetic flux, plage and sunspot filling factors are plotted alongside coeval TSI and S-index observations in Figure~\ref{fig_all}. 
The SDO/HMI disc-averaged quantities shown in Figure~\ref{fig_all} are at the maximum cadence considered in this study (1 observation every 4 hours, \ie~6 per day).
We then average the SDO/HMI quantities over daily bins and concatenate these timeseries with the timeseries of S-index and TSI. We show the RV variations as a function of unsigned magnetic flux, S-index and TSI in the top row of Figure~\ref{fig_correlations}. The bottom row shows the three activity indicators plotted as a function of each other.
Figure~\ref{fig_correlations} shows that the RV variations correlate much better with the unsigned magnetic flux (R = 0.92) than the S-index (R = 0.75) or optical photometry (R = 0.46).
Observations of high sunspot coverage are highlighted in yellow. We see that the Sun is dominated by spots at high activity levels.
At the peak of the activity cycle, the Sun's photometric variations are anti-correlated with \ca~variations \citep[][Fig.1]{radick2018}, so the Sun is spot-dominated; at lower activity levels, they are positively correlated, implying that the solar surface is dominated by faculae/plage. This is in agreement with previous studies \citep{frohlich1998,krivova2007,shapiro2014}.

Following \citet{meunier2019c} who reported on a hysteresis between \drv~and \ca~in solar Cycle 23, we investigate the hystereses between \drv, S-index, and \bhat in Cycle 24. We observe hystereses between all quantities (shown in Figure~\ref{fig_hysteresis}) and discuss their physical origins in detail in Appendix~\ref{app_hysteresis}.

Time lags of 1-3 days between RV variations and the bisector span and FWHM have been reported previously in spectroscopic HARPS-N observations of the Sun \citep[][Fig.15]{colliercameron}.
We cross-correlate \drv, \bhat, and the S-index against each other to look for time shifts between them. 
We do not find any significant time shifts between any of our observables. 

\subsection{Periodogram analysis}
\begin{figure*}[t]
\centering
\includegraphics[scale=0.45]{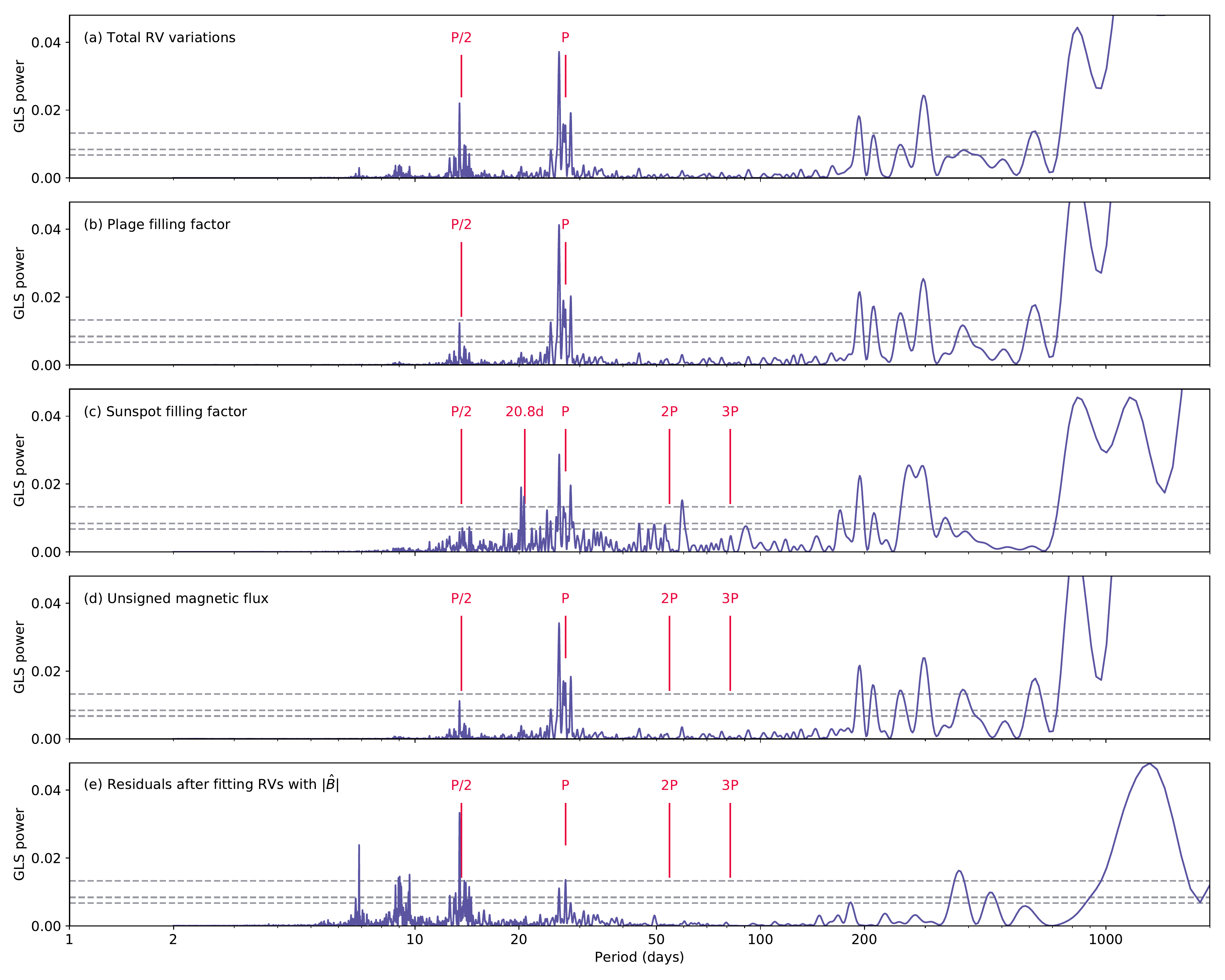}
\caption{Generalised Lomb-Scargle periodograms of (a) the total SDO/HMI-derived RV variations, (b) the filling factor of plage, (c) the filling factor of sunspots, (d) the unsigned magnetic flux \bhat, and (e) residuals after modelling \drv~with a linear fit of \bhat. The horizontal grey dashed lines represent the false alarm probability levels (from bottom to top: 10\%, 1\% and 0.001\%).  \label{fig_GLS}}
\end{figure*}

Figure~\ref{fig_GLS} shows Generalised Lomb-Scargle periodograms \citep{Zechmeister2009} of the RV variations (panel (a)), the filling factors of plage (panel (b)) and sunspots (panel (c)) and the unsigned magnetic flux (panel (d)).
Most of the periodicity below 100 days is confined to periods close to the rotation period and its first harmonic. It is worth noting that these peaks are in fact forests of peaks, in which several peaks are significant above the 0.001\% confidence level. This means that depending on when or for how long we might observe the Sun, we may measure rotation periods differing by several days \citep[\eg][]{mortier2017,nava2019}.
In the periodogram of the sunspot filling factor, we also detect a significant peak consistent with the 20.8 day peak detected by \citet{Lagrange2010I}. This peak is possibly related to the lifetime of the spots. Alternatively, it may be associated with global-scale equatorial Rossby waves (r-modes) that produce oscillations on a 19-day recurrence timescale \citep{lanza2019}. In this periodogram, we see many significant peaks around the rotation period and its harmonics. Additionally, some peaks are significantly different from the rotation period or its harmonics; as \citet{Lagrange2010I} previously emphasized, we should be careful when attributing these signals to non-activity processes. The strong signals in the spot filling factor do not necessarily translate into signals in the total RV variations, because the Sun is faculae-dominated for the majority of its cycle \citep[\eg][]{frohlich1998,krivova2007,shapiro2014}. However, in younger, faster rotating Sun-like stars whose behaviour has been observed to be spot dominated \citep[see][]{Lockwood2007,radick2018}, we would certainly expect the RV variations to show a more ``spot-like" periodogram structure.

\section{Modelling $\Delta$~RV using \bhat}
\label{sec_model_b}
\label{sec_active}

\begin{figure*}[t]
\centering
\includegraphics[scale=0.8]{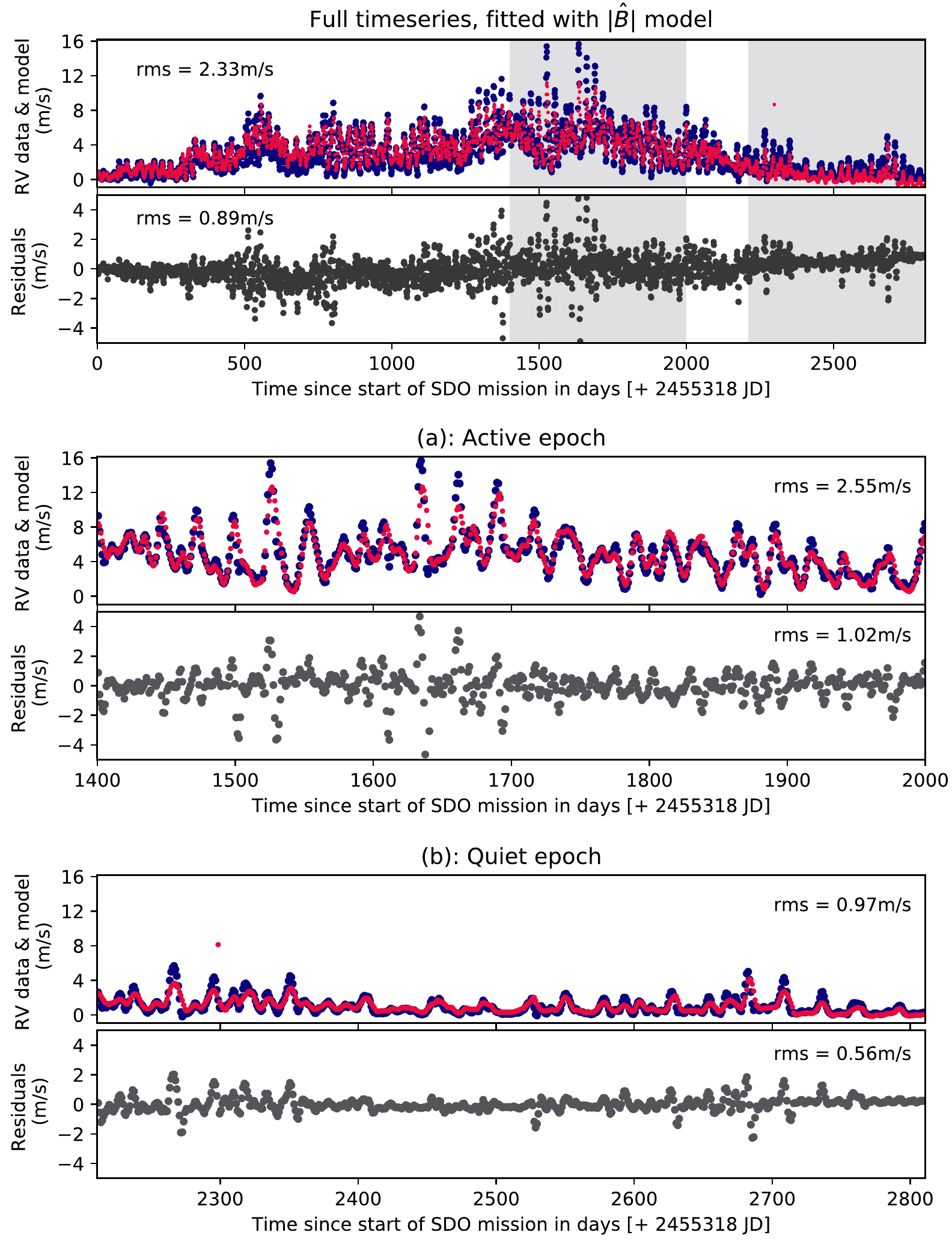}
\caption{Top panel: Estimated \drv~(in blue) for the full (non-concatenated) SDO/HMI RV dataset, modelled with a linear fit in unsigned magnetic flux (red);  residuals of the fit (grey). Panel (a): same as top, but for the 600-day epoch of high magnetic activity. Panel (b): same as top, but for the ``quiet" 600-day epoch of low magnetic activity.
\label{fig_model_b}}
\end{figure*}

We model the RV variations estimated in Section~\ref{sec_reconstruct}, \drv(t) as a linear model of \bhat:
\begin{equation}
    \Delta~RV(t) = \alpha \frac{|\hat{B}_{\rm obs}|(t)}{\langle|\hat{B}_{\rm obs}| \rangle} + RV_0
\end{equation}
where $\alpha$ is a constant scaling factor and $RV_0$ is a constant zero-point offset. $\langle |\hat{B}_{\rm obs}|\rangle $ is the mean of \bhat~over the full timeseries. 
We optimise the parameters $\alpha$ and $RV_0$ via a least squares procedure.
We model the full \drv(t) timeseries of daily averages (plotted in the top panel of Figure~\ref{fig_all}). 
The fit over the full magnetic cycle is shown in the top panel of Figure~\ref{fig_model_b}. 
To examine the performance of the linear \bhat~model as a function of magnetic activity levels, we model two separate stretches, at activity maximum and minimum, spanning 600 days each. 
The low-activity, quiet epoch ranges between 2016 May 20 (JD~=~2457529) to 2018 January 10 (JD~=~2458129), which corresponds to days 2211--2811 in the figures.
The high-activity epoch spans 600 days from 2014 March 1 (JD~=~2456718) to 2015 October 22  (JD~=~2457318; days 1400--2000 in the figures).
The ``active'' and ``quiet`` fits are plotted in panels (a) and (b) of Figure~\ref{fig_model_b}.
The estimated parameters for all fits are reported in Table~\ref{table_results}.
The root mean scatter (\rms) of the full dataset is 2.33~\ms, and that of the residuals is 0.89~\ms.
Overall, a simple \bhat~model reduces the \rms~of the RV variations by 62\%, \ie~a factor of 2.6.
Although we do see correlated residuals at times of high activity, the residuals over the full cycle are flat.

\paragraph{Rotationally modulated RV variations at times of low activity}

The dominant process at play is the suppression of convective blueshift incurred by areas of plage \citep{milbourne}. \bhat~correlates well with their presence, as seen in Figure~\ref{fig_ffactors} in Appendix~\ref{app_correlations}. There are very few spots; the maximum filling factor of sunspots in this 600-day stretch is 0.03\% (compared with 0.14\% in the active stretch and 0.09\% overall). We therefore expect \bhat~to correlate well with RV variations.
Indeed, this model is an excellent fit during activity minimum, as evidenced in Figure~\ref{fig_model_b}b. 
Over the quiet epoch, the RV residuals have an \rms~of 0.56~\ms.

\paragraph{Rotationally modulated RV variations at times of high activity}

\begin{figure*}[]
\centering
\includegraphics[scale=0.58]{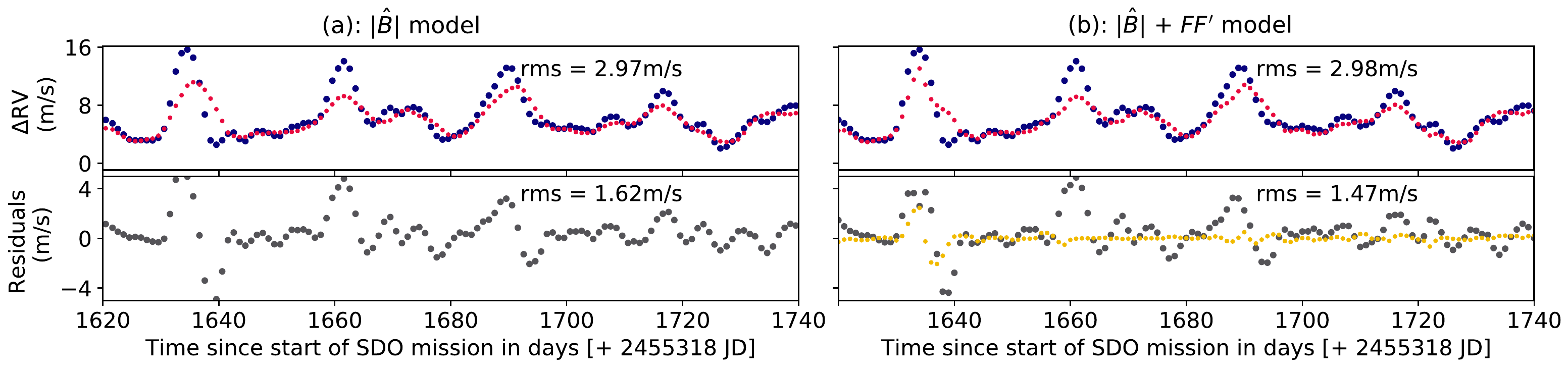}
\caption{Zoom-in on \drv~over 4 to 5 solar rotations (120 days) at activity maximum, to highlight the smooth, rotationally modulated signals in both the observations (blue) and the residuals (grey) after fitting the model (red). Panel (a): \bhat~model of Section~\ref{sec_model_b}. Panel (b): \bhat~+~\ff~model of Section~\ref{sec_model_bff}; the \ff~term is shown (yellow) in the residuals panel.
\label{fig_zoom}}
\end{figure*}

The best-fit estimates of the model parameters ($\alpha$ and $RV_0$) differ significantly from the fit at low magnetic activity (see Table~\ref{table_results}). 
This is because convective blueshift is more suppressed by larger magnetic structures (which are more prevalent in periods of high activity), \ie~the faculae in concentrated areas of plage, as previously found by \citet[][Fig.7]{meunier2010b} and \citet[][Fig.6]{milbourne}. 
The \bhat~model accounts for RV variations down to a residual \rms~of 1.02~\ms. As seen in Figure~\ref{fig_model_b}~(a), significant rotationally modulated RV variations remain unaccounted for. 
In Figure~\ref{fig_zoom}(a), we zoom-in further on the RV variations over 4 to 5 solar rotations at activity maximum. This stretch spans 120 days from 2014 October 6 (JD = 2456937) to 2015 February 3 (JD~=~2457057; days 1620--1740 in the figures). The smooth, rotationally modulated signal that remains in the RV residuals is very clear on this timescale.
At activity maximum, we expect suppression of convective blueshift to produce significant RV variations \citep{meunier2010a,meunier2010b,haywood2016}. Additionally, there are sunspots, which now produce significant RV variations by blocking Doppler-shifted flux on the rotating solar surface \citep{saar1997,Lagrange2010I,meunier2010b,haywood2016}.
The relationship between this sunspot flux-blocking RV term and \bhat~is more complex than for the RV due to suppression of convective blueshift. In fact, they do not correlate with each other (R = 0.08). When a spot crosses the central meridian, this RV contribution is zero, while \bhat~would be at its maximum. Therefore, a simple, linear \bhat~model cannot adequately capture RV variations from sunspot flux-blocking.

To investigate this hypothesis, we apply the technique developed by \cite{Aigrain:2012}. Their \ff~term accounts for RV variations incurred by brightness inhomogeneities on a rotating disk.

\begin{table*}[]
\centering

\begin{tabular}{cccccccc}
\centering
{\bf Model} & 
{\bf Span of data} &
\multicolumn{3}{c}{\bf Parameter estimates}	&
\multicolumn{2}{c}{\bf RMS values}		&
{\bf \% reduction}

\\   
                                      
    &
{\bf modelled}  & 
\multicolumn{3}{c}{\bf (\ms)} & 
\multicolumn{2}{c}{\bf (\ms)} &
 {\bf in RV RMS}  \\

&
&
$\alpha$  &
$\beta$ &
RV$_{0}$ 	&
Data        &
Residuals     &
{\bf amplitude}        \\ \\
\hline                                                                                          
\\

\bhat~model   &  
Full timeseries	& 9.616$\pm$~0.008 &  -  &	-6.777$\pm$~0.008 &	2.33 &	0.89 &	62\%   \\                   

\bhat~model  &                                       
Quiet epoch & 8.37$\pm$~0.04 &  -  &	-5.32$\pm$~0.03 & 	0.97 &	0.56	& 43\%   \\

\bhat~model    &                                                 
Active epoch &	11.49$\pm$~0.02   &	- & -8.81$\pm$~0.02 &	2.55 &	1.02 &	60\% \\

\bhat~with \ff~model   &  
Full timeseries	& 9.366$\pm$~0.009 &  -1.99$\pm$~0.02  &	-6.678$\pm$~0.009 &	 2.27 &	 0.85 &	63\%   \\      

\bhat~with \ff~model   &                                     
Quiet epoch & 8.40$\pm$~0.04 &	 -1.81$\pm$~0.04   & -5.44$\pm$~0.03 &	0.98 &	0.51 &	47\%   \\  

\bhat~with \ff~model   &                                                 
Active epoch &	11.24$\pm$~0.02 &  -2.48$\pm$~0.03  &	-8.72$\pm$~0.02 &	 2.55 &	0.97 &	62\%   \\ 

\\

\hline                                         

\end{tabular}

\caption{Best-fit parameter estimates and \rms~values for each of the models tested in Sections~\ref{sec_model_b} and~\ref{sec_model_bff}. 
The quiet and active epochs span 600 days each.
Note that the \bhat~+~\ff~model can only be run on days for which both SDO/HMI and SORCE TSI data are available (hence the slightly different data \rms~estimates).}

\label{table_results}

\end{table*}

\section{Modelling \drv~using both \bhat~and the \ff~model}
\label{sec_model_bff}

\begin{figure*}[t]
\centering
\includegraphics[scale=0.8]{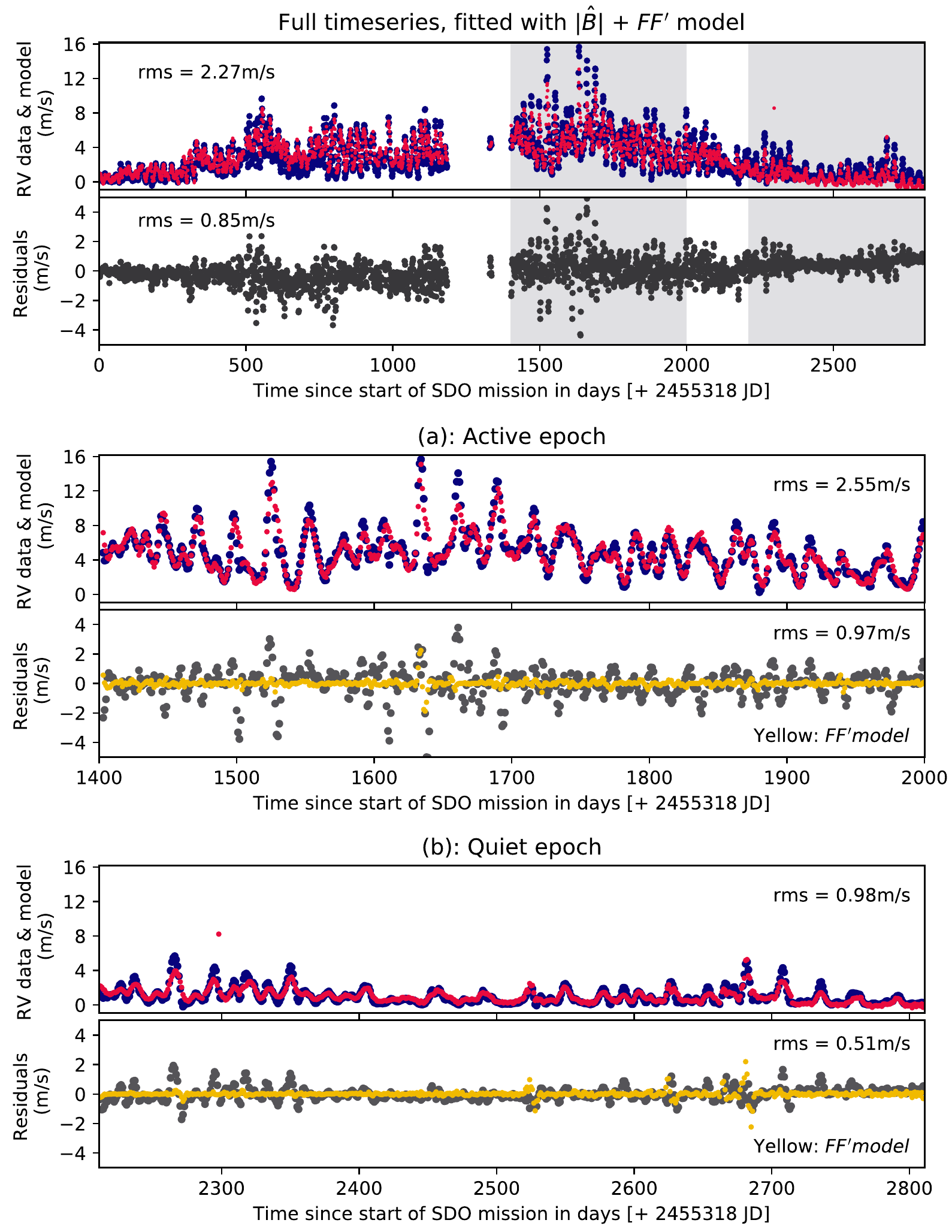}
\caption{
Top panel: Estimated \drv~(in blue) for days where TSI measurements are available from SORCE, modelled with a linear fit in unsigned magnetic flux and the \ff~term (red);  residuals of the fit (grey). Panel (a): same as top, but for the 600-day epoch of high magnetic activity. Panel (b): same as top, but for the ``quiet" 600-day epoch of low magnetic activity. For comparison, the \ff~term is overplotted (yellow) alongside the residuals in panels (a) and (b).
\label{fig_model_bff}}
\end{figure*}

A simple \bhat~model fits RV variations well overall, but becomes insufficient at the peak of the solar magnetic cycle (see Section~\ref{sec_active}), where significant, correlated residuals remain, as visible in Figure~\ref{fig_model_b}~(a).
We apply the method  of \citet{Aigrain:2012} in an attempt to account for RV variations from flux-blocking from sunspots (and plage). 

As derived in \citet[][Eqn.10]{Aigrain:2012}, the RV perturbation due to a spot crossing the disc can be expressed as follows: 
\begin{equation}
\Delta RV_{\mathrm{rot}}(t) = - \frac{\dot{\Psi}(t)}{\Psi_{0}} \, \Bigl [ 1- \frac{\Psi (t)}{\Psi_{0}} \Bigr ] \, \frac{R_{\star}}{f_{\rm spot}},
\end{equation}
where $\Psi(t)$ is the observed solar flux, $\Psi_{0}$ is the solar flux for a non-spotted photosphere and $\dot{\Psi}(t)$ is the first time derivative of $\Psi(t)$. $R_{\star}$ is the solar radius. The parameter $f_{\rm spot}$ represents the drop in flux produced by a spot at the centre of the solar disc, and corresponds to the sunspot filling factor. 
Since both dark inhomogeneities (from spots) and bright inhomogeneities (from faculae, mainly in plage) produce RV perturbations \citep[\eg][Fig.7]{meunier2010a}, we use the total magnetic filling factor $f_{\rm all}$ rather than $f_{\rm spot}$ alone. 
We find that when using $f_{\rm spot}$, the \ff~term has an \rms~amplitude of 0.02~\ms, \ie~10 times less than when we consider all magnetic elements, including plage. 
We write the following formulation:
\begin{equation}
    \Delta RV(t) = - \frac{F'(t)}{\langle F' \rangle} (1 - \frac{F(t)}{\langle F \rangle}) \frac{1}{f_{\rm all}(t)}
\end{equation}
where $F$ and $F'$ correspond to the TSI (Section~\ref{sec_tsi}) and its first time derivative, respectively. $\langle F \rangle $ and $\langle F' \rangle $ are the means of the TSI lightcurve and its first time derivative, respectively.  
To compute $F'$, we interpolate the TSI observations ($F$) onto an evenly, over-sampled array and then fit them using Gaussian-process regression using a basic square exponential kernel \citep{Rasmussen2006}. We then compute the derivative using second-order accurate central differences.
We multiply the \ff~term above by a normalising factor $\langle f_{\rm all} \rangle{\langle F \rangle}/{\langle F' \rangle}$ so that the full term is of order unity.
Our resulting \bhat~+~\ff~model is as follows:
\begin{equation}
    \Delta~RV(t) = \alpha \frac{|\hat{B}_{\rm obs}|(t)}{\langle |\hat{B}_{\rm obs}| \rangle} 
    - \beta \frac{F'(t)}{\langle F' \rangle} (1 - \frac{F(t)}{\langle F \rangle}) 
    \frac{\langle f_{\rm all} \rangle}{f_{\rm all}(t)}
    + RV_0 
\end{equation}
where $\beta$ is a scaling factor that we fit for in our least-squares optimisation, along with $\alpha$ and $RV_0$. 
The resulting best fit when modelling the full timeseries (where the SDO/HMI and SORCE data overlap) is shown in the top panel of Figure~\ref{fig_model_bff}.
We also apply this model to the active and quiet epochs, as shown in panels (a) and (b) of Figure~\ref{fig_model_bff}.
The best-fit estimate of $\beta$ is non-zero (see Table~\ref{table_results}), and the \ff~term has an \rms~of order 0.2~\ms.
For the full timeseries, the \rms~of the residuals (0.85~\ms) is slightly lower than that obtained with the \bhat~model (0.89~\ms). 
However, the overall fits with \bhat~+~\ff~are very similar to those resulting from modelling the RVs with \bhat~only (as is done in Section~\ref{sec_model_b}).
The residuals still display correlated behaviour. Clearly, the \ff~model does not fully account for the signals leftover from the \bhat~model.

\section{Rotationally modulated RV variations not traced by \bhat~or \ff} \label{sec_others}

The residuals from both models tested (\bhat~ and \bhat+\ff) display correlated behaviour during times of high magnetic activity. We propose two explanations.

\label{sec_beyond_ff}
First, the \ff~term does not adequately fit the RV variations resulting from sunspot flux-blocking (Section~\ref{sec_model_bff}).
Several studies have previously found that the \ff~method cannot fully account for RV variations \citep[\eg][]{Oshagh17,Bastien14,Haywood2014}.  
We note that the \ff~term (or the $F^2$ term that can be used to account for suppression of convective blueshift) is not {\it expected} to match RV variations perfectly, because $F'$ includes the derivative of the limb darkening of $F$, which should not be a part of the RV model. We demonstrate this in detail in Appendix~\ref{app_ffmethod}.

\label{sec_beyond_physics}
Another possible explanation for the RV residuals is that there are additional processes at play, which are either missing from the RV model of \citet{milbourne} and \citet{haywood2016} used to estimate RV variations, or that do not correlate directly with \bhat~(or \ff).
This finding is consistent with that of \citet{miklos2019}, who investigated the activity-sensitivity of spectral lines observed by HARPS-N and concluded that there must be additional factors, not yet accounted for by current state-of-the-art models such as \citet{meunier2017}.
Other types of surface velocity fields not included in our model may give rise to rotationally modulated RV variations, such as:

    \paragraph{Evershed flows} Sunspots are made of umbral and penumbral regions. Evershed flows, which are contained within penumbral regions and flow radially outward from the central umbra to the outer edge of the sunspot, are tangential to the surface \citep{evershed1909}. They will be most visible in RV for sunspots located away from disc centre, where their flows are more directed along our line of sight \citep[\eg][Fig.6]{haywood2016}.
    
    \paragraph{Moat flows} Outside of the penumbra, but within the active region, are so-called moat flows (\eg, \citet{Solanki2003} and references therein). These are also tangential to the surface but weaker than penumbral flows (of order 1/4 to 1/2 as strong), but as they are brighter (since they are typically in plage) and normally cover an area $\approx 3\times$ larger than the spot. They may contribute significantly to the total RV signal \citep[\eg][]{iampietro2019}.
    The effect of moat and Evershed flows on the disc-averaged solar RV variations is being investigated by \citet{Saar2020}.
    
    \paragraph{Active region inflows} \citet{Gizon2001} report the presence of inflows towards active regions on the Sun's surface, with amplitudes up to 50~\ms~\citep[see][Sect.7.1.]{Gizon2010}.  These are also currently under study \citep{Saar2020}.

    \paragraph{Unresolved flows} We could be seeing residual effects arising from unresolved flow motions and magnetic processes taking place within magnetically active SDO pixels. HMI samples the Fe~{\sc I} line profile (at 6173~\AA) at only six points in wavelength (cf. Section~\ref{sec_sdodata}). The line shift (velocity), depth, width, magnetic field and continuum intensity are then determined by fitting a symmetric Gaussian to these points. 
    This coarse sampling at the pixel level, and the fact that pixels are of a size comparable to that of granules mean that we are missing spectral-line asymmetries from processes taking place below the pixel resolution, \eg~due to convection. \citet{Saar09} makes some initial attempts \citep[expanding on][]{Saar03} at including convective line asymmetries in plage models.   
    
    \paragraph{Zeeman broadening}
A final item not included in our model is the direct effect of magnetic fields on the line profiles themselves \citep{reiners2013}. 
Due to Zeeman broadening in magnetic regions, the line profiles originating there are wider and shaped differently. If the lines are stronger, they can show enhanced equivalent widths
as well. These differences lead to subtle RV changes as active regions rotate and change in number and size.
\citet[][]{reiners2013} showed that RV amplitudes resulting from Zeeman broadening were of the order $\Delta v_B \approx$ 300 $f (B/ 1 $kG$)^2 (\lambda/1 \mu$m)$^2$ \ms, where $f$ is the magnetic filling
factor, $B$ is the local magnetic field strength, and $\lambda$ is the wavelength. Note that HMI measures fluxes (\ie~$B$ times area) and the actual $B$ in resolved plage flux tubes is $\sim$ 1.5 kG  \citep[\eg][]{Buehler15}. Most plage contains a mix of fluxtubes and field-free areas. Thus the true solar magnetic $f$ is close to 1-2 \% (note that we are ignoring the ubiquitous weak turbulent fields in this estimate). Adopting $\langle \lambda \rangle = 0.5 \mu$m, we can estimate $\Delta v_B \approx$ 1.7 -- 3.4 m s$^{-1}$.  Note that Zeeman broadening is only partly and imperfectly removed by the fit to $B$ since the actual RV dependence is $\propto f \lambda^2 B^2$. Since proper treatment would require a different calculation of the filling factors, and due to the wavelength dependence, a different computation of RVs, we leave exploring this to a future paper.  We note, however, that the Zeeman broadening effect would follow $B$ in phase, and could reduce residuals coincident with large
$B$ concentrations during active epochs.

\section{Using \bhat~to confirm and characterise long-period, low-mass planets}
\label{sec_planet}

Previous studies have shown that we \emph{must} account for RV variability in order to detect and characterise low-mass, long-period planets \citep[\eg][and others]{Saar09,hall2018,meunier2019b}.
Here, we test whether the unsigned magnetic flux could, in principle, be used to mitigate rotationally modulated RV variations to characterise small planets accurately and precisely \citep[\eg~to better than 10\% precision in mass;][]{zeng2013}.
The present analysis is not intended to be a comprehensive exploration of parameter space, nor is it meant to reflect realistic ground-based observing conditions for stellar RV surveys.
When facing reality, the most important factor to consider will be the precision to which \bhat~can be measured; we discuss this in Section~\ref{sec_prospects_b}.
The effects of magnetoconvection, \ie~granulation and supergranulation will also need to be accounted for as they produce RV variability at the \ms~level \citep[\eg][]{meunier2015}.

\subsection{Procedure}
Here, we inject synthetic planet signals in the 8-year long, daily averaged SDO/HMI-derived \drv~timeseries (see Figure~\ref{fig_all}), down-sampled to 6 months per year, to simulate the visibility pattern of a star observed from the ground \citep[\eg][Sect.3.3]{hall2018}. For simplicity, we do not remove observations to mimic weather losses, as \citet{hall2018} estimate that targets visible during the summer only experience a 6\% loss of nights due to weather at optimal observing sites. We are left with 1456 daily observations spread over 8 consecutive seasons.
We consider three RV semi-amplitudes $K$: 0.5, 0.3 and 0.1~\ms. In each case, we consider a circular orbit, with period $P_{\rm orb}$~=~\Ppl~days and time of transit $t_0~=~1501.92+2455318$~days (close to the mid-time of the datatset). We choose $P_{\rm orb}$ close to but not exactly at 300 days to avoid overlap with any potential aliases arising from SDO's geosynchronous orbit.
Assuming an orbital inclination of 90 degrees and a 1~\msun~host star with a typical uncertainty for a bright, solar analogue of 0.03~\msun, an RV semi-amplitude 
of 0.5~\ms~corresponds to a planet with a mass of 5.2~\mearth, 0.3~\ms~corresponds to 3.1~\mearth, 
and 0.1~\ms~corresponds to 1.04~\mearth.

For each scenario, we fit models consisting of a circular Keplerian and zero-point offset, and either the linear function of \bhat~(of  Section~\ref{sec_model_b}), or the \bhat~+ \ff~combination (of Section~\ref{sec_model_bff}).
We account for the remaining residuals by adding the residual RMS values (0.89 and 0.87~\ms~for the \bhat~and \bhat~+\ff~models, respectively; see Table~\ref{table_results}) in quadrature to the 1-sigma RV uncertainties (0.1~\ms, see Table~\ref{table_budget}). Rounding up, we obtain an effective RV uncertainty of 0.9~\ms~for both models. 
In a real-case scenario one would implement a correlated noise framework (\eg~Gaussian process regression) to ensure that the parameter estimates are as accurate and precise as they can be in the presence of correlated noise. This statistically demanding analysis is beyond the scope of the present analysis, whose primary purpose is to determine whether the planets can be recovered.
We assume prior indication of a planet in this range of orbits, \eg~through the detection of one or more transits, so we impose broad Gaussian priors of 10 and 20 days on $P_{\rm orb}$ and $t_0$ respectively.
We maximise the likelihood of each model and determine the best-fit parameter values through an MCMC procedure similar to the one described in \cite{Haywood2014}, in an affine-invariant framework \citep{goodman2010}. 
In all cases, the MCMCs reveal a parameter space with multiple local maxima in likelihood.
As is done routinely in exoplanet analyses, if the majority of the MCMC chains give the same solution, we remove deviant chains before estimating the model parameters and uncertainties. However, if no single area of parameter space is clearly preferred, we deem the MCMC outcome as a non-detection.

\subsection{Outcomes}
\begin{table*}[]
\centering

\begin{tabular}{lccccccccc}
\centering
{\bf Activity} & 
\multicolumn{3}{l}{\bf Injected planet parameters} &
\multicolumn{6}{l}{\bf Best-fit parameter estimates and 1-$\sigma$ uncertainties}	
\\                                                                             

{\bf model}&
$K$ &
$P_{\rm orb}$ &
$t_{0}$ &

$\alpha$  &
$\beta$ &
$RV_{0}$ 	&
$K$ &
$P_{\rm orb}$ &
$t_{0}$ 
\\

&
(\ms)     &
(days)    &
(days)    &

(\ms)     &
(\ms)     &
(\ms)     &
(\ms)     &
(days)    &
(days)  
\\ \\
\hline                                                                             
\\

\bhat   &  
0.5 &
300.38  &
1501.92 &
9.5~$\pm$~0.1 &  -  &	-6.8~$\pm$~0.1 &
0.48~$\pm$~0.03   &   297~$\pm$~1  & 1512~$\pm$~3
\\

\bhat~+~\ff   &  
0.5 &
300.38  &
1501.92 &
9.5~$\pm$~0.1 &  -2.5~$\pm$~0.2  &	-6.8~$\pm$~0.1 &
0.47~$\pm$~0.03   &   297~$\pm$~1  & 1515~$\pm$~3
\\

\bhat   &  
0.3 &
300.38  &
1501.92 &
9.6~$\pm$~0.1 &  -  &	-6.8~$\pm$~0.1 &
0.29~$\pm$~0.03   &   296~$\pm$~1  & 1520~$\pm$~5
\\

\bhat~+~\ff   &  
0.3 &
300.38  &
1501.92 &
9.5~$\pm$~0.1 &  -2.5~$\pm$~0.2  &	-6.8~$\pm$~0.1 &
0.29~$\pm$~0.03   &   295~$\pm$~2  & 1526~$\pm$~6
\\

\bhat   &  
0.1 &
300.38  &
1501.92 &
9.8~$\pm$~0.1 &  -  &	-7.0~$\pm$~0.1 &
0.32~$\pm$~0.04   &   335~$\pm$~2  & 1397~$\pm$~6
\\

\bhat~+~\ff   &  
0.1 &
300.38  &
1501.92 &
9.7~$\pm$~0.1 &  -2.4~$\pm$~0.3  &	-7.0~$\pm$~0.1 &
0.24~$^{+0.04}_{-0.06}$   &   334$^{+3}_{-42}$  & 1394~$^{+152}_{-8}$
\\

\\

\hline                                                                                                            
\end{tabular}

\caption{Best-fit parameter estimates for each of the planet injection tests made in Section~\ref{sec_planet}. 
Note that 0.5~\ms~and 0.3~\ms~injected planets are recovered while the 0.1~\ms~injected planet is not recovered (see Section~\ref{sec_planet} for details).
The times of transit $t_0$ have 2455318~JD subtracted from them.}

\label{table_planets}

\end{table*}

The results of all scenarios are presented in Table~\ref{table_planets}.
In all cases, the estimates for $\alpha$, $\beta$ and $RV_0$ match those determined via the least squares optimisation procedures of Sections~\ref{sec_model_b} and~\ref{sec_model_bff} within 1 to 2-$\sigma$ (see Table~\ref{table_results}).

\begin{figure}[t]
\centering
\includegraphics[scale=0.43]{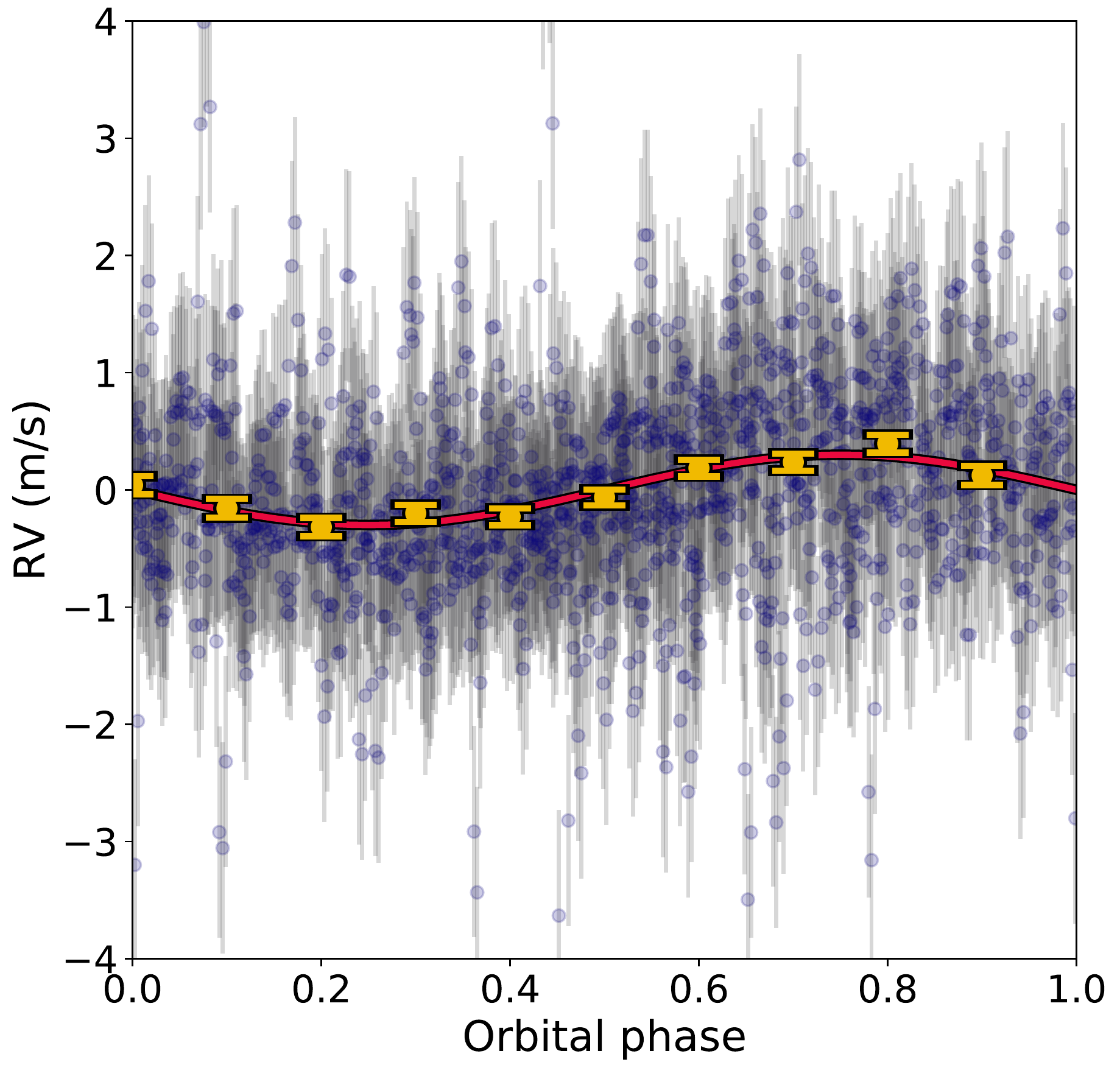}
\caption{Phase plot of the orbit of an injected planet with K~=~0.3~\ms, at P$\simeq$~300~days (red line) recovered when modelling the RV observations with a linear fit of \bhat. Blue points: RV observations after subtracting the \bhat~model and zero-point RV offset. Yellow points: inverse-variance weighted averages of the blue points in bins of size 0.1 in phase. 
The fits work well down to K~=~0.3~\ms but fail at 0.1~\ms (see Section~\ref{sec_planet} for details).
\label{fig_phase}}
\end{figure}

\paragraph{Injected planets with $K$~=~0.5 and 0.3~\ms}
We recover the planet signal for both of these $K$ amplitudes. The RV amplitudes are estimated accurately within 1-$\sigma$. The orbital periods are systematically underestimated by up to 3-$\sigma$, as was found by \citet[][Fig.5]{hall2018} who performed very similar simulations. 
The orbital phases $t_0$, too, are off from their correct value by up to 3-$\sigma$, due to $P_{\rm orb}$ being offset.
The planet amplitude is recovered with the same significance using either the \bhat~model or the \bhat~and \ff~combination.
We show the phase-folded orbit of the injected $K$~=~0.3~\ms, recovered using the \bhat~model in Figure~\ref{fig_phase}.

\paragraph{Injected planet $K$~=~0.1~\ms}
Neither activity models are sufficent to recover signals of this amplitude.
The parameter space explored by the MCMC chains shows multiple solutions with similar likelihoods at the 300-day period of the injected planet, as well as at 330 days and near 400 days.
When we inject a planet at 300 days with $K$~=~0.1~\ms, we systematically find that the most likely solution is for a signal with a period of 335 days with an RV amplitude of 0.3~\ms.
The 335-day signal does not appear to be caused by uneven sampling, as we see complete and uniform coverage at all phases.

\paragraph{Stellar activity signals in the 300-day range}
We see several significant peaks at 200-400 days in the periodogram of the RVs~(Figure~\ref{fig_GLS}a). \bhat, too, exhibits significant and similar (but non-identical) periodicities in the same range (Figure~\ref{fig_GLS}d). Panel (e) shows that the RV residuals (after applying the \bhat~model) exhibit comparatively less power in this period range, but we still see two significant peaks in the 350-500 day range.
\citet[][Fig.10]{meunier2010b} also detect significant peaks at 300-400 days in solar RV variations of Cycle 23. Their RV variations are estimated using an independent method, using catalogues of sunspot and plage records and magnetograms from SoHo/MDI (which is in a different orbit than SDO).
The most likely explanation for the nature of the 335-day signal is that it is a long-term signature of magnetic activity. 

\subsection{From this idealised scenario to stellar observations}\label{sec_stars}
In this idealised setup, we successfully retrieve 300-day planet orbits with RV amplitudes down to 0.3~\ms, while 0.1~\ms~signals remain out of reach.  
To break the 0.1~\ms~barrier, additional RV signals not well traced by either either \bhat~or \ff~will need to be modelled adequately. 
More generally, these planet injection tests show that in order to access planets with periods of a few hundred days, we will need to model all stellar signals that have similar periods and amplitudes much larger than $K$.
The \bhat~ and the combined \bhat~+~\ff~models perform equally well at this orbital period range.
This is expected since we obtain very similar RV residuals with both models (see Sections~\ref{sec_model_b} and~\ref{sec_model_bff}).
The planet retrievals carried out here are a best-case scenario that only considers \emph{rotationally modulated} RV variations from magnetically active regions. 
In stellar observations, there will be additional intrinsic variability from magnetoconvection. For this dataset, it would have an \rms~of 1.1~\ms~given our cadence and sampling strategy \citep{meunier2015}.
Long-baseline stellar observations are expected to feature RV variations from large-scale meridional circulation, which varies with the magnetic cycle \citep{komm1993,meunier1999,ulrich2010}. Although meridional flows have not yet been clearly identified in stars other than the Sun, they produce peak-to-peak RV amplitudes of 1-1.4~\ms~when viewing the Sun edge-on, and 2.3-3.3~\ms~in a pole-on scenario \citep{meunier2020}. \citet{meunier2020} extend their results to other Sun-like stars and predict peak-to-peak amplitudes of up to 4~\ms~in stars with strong magnetic cycles seen pole-on.
The significance of a planet detection will depend strongly on how precisely we can measure the unsigned magnetic flux in Sun-like stars; see further discussion in Section~\ref{sec_prospects_b}.
Additionally, there will be instrumental systematics of order 0.1-1~\ms~for current-generation spectrographs \citep{fischer2016}. In particular, wavelength calibration and long-term stability remain challenging even in current state-of-the-art spectrographs \citep[\eg][]{Cosentino2012}. 
Our RV and \bhat~timeseries are sampled simultaneously, and every night for 8 seasons; ground-based surveys will suffer losses from poor weather \citep[\eg][]{hall2018}. These caveats will impact the performance of the technique presented here and diminish the significance of the mass determinations.
We note that a systematic investigation of the detectability of low-mass, long-period planets is under way \citep{langellier2020}.
They explore a broad range of parameter space using HARPS-N RV observations of the Sun, so their analysis is highly complementary to the one here. \citet{langellier2020} inject a wide variety of planet signals into solar observations and recover them by treating magnetic activity using Gaussian process regression. 
For an in-depth investigation of the impact of observation strategies on the detectability of Earth-mass, long-period planets, we refer the reader to \citet{hall2018}.

\section{Future prospects}\label{sec_prospects}

\subsection{Prospects for measuring \bhat~ for other stars}
\label{sec_prospects_b}
The significance of an exoplanet detection will depend strongly on how precisely we can measure the unsigned magnetic flux in stars other than the Sun. 
Zeeman Doppler imaging \citep[ZDI;][]{donati1997b} has long been used to image the large-scale (polarised) magnetic field structures of stars, particularly those that are fast rotating and much more active than the Sun (see \citet{reiners2012} and references therein). However, RV variations stem from magnetic fields taking place on much smaller spatial scales than those probed by ZDI.
In principle, it is possible to measure small-scale, unsigned magnetic flux by examining Zeeman broadening in magnetically sensitive spectral lines of stars \citep{robinson1980,saar1988}. 
It has been detected, for example, in the younger, moderately active, faster rotating ($P_{\rm rot} \approx$ 12 d) K2 dwarf Epsilon Eridani, which 
has an average unsigned magnetic flux in the range 125 - 200 G \citep{valenti1995,lehmann2015}.
The Sun's average unsigned magnetic field is twenty times smaller (10~G).
Several studies have attempted to measure Zeeman broadening in slowly rotating, relatively quiet late-type stars  \citep[\eg][]{saarlinsky1986,saar1986,basri1988,rueedi1997,Anderson10}. More recently, \citet{Kochukhov20} developed a method employing multiple lines with different Zeeman splitting patterns and making use of differential magnetic intensification of line equivalent widths \citep{Basri92,Saar92}. This work looks very promising for accurately pushing \bhat~detections to lower levels: using only eight spectral lines, \citet{Kochukhov20} detect filling factors as low as $f~=$~7\% and unsigned average fields as low as 220 G.  Modeling more lines should yield further improvements in determining \bhat~at low levels.  
\citet{mortier2016} combine multiprofile  least squares deconvolution \citep[originally proposed by][]{kochukhov2010} with singular value decomposition to extract the unsigned magnetic flux from thousands of spectral lines. Their preliminary application to high resolution spectra from HARPS and HARPS-N of an inactive K3 dwarf gives encouraging results.

The aforementioned studies identify several avenues to improve our prospects of detecting Zeeman broadening in old, quiet Sun-like stars. These include: improving  modelling of spectral lines (better atomic data and better understanding of line broadening), particularly in the (near-)IR where Zeeman broadening is stronger as it has a squared dependence on wavelength; improving our constraints on convection and turbulence in the stellar atmosphere; and improving our understanding of the impact of line blending and telluric contamination. 
The magnetic-region filling factor and magnetic field strength are, to some extent, degenerate, but their product, $B.f$ is more easily measurable \citep[\eg][and references therein]{gray1984,saar1988,reiners2012}.
For example, the SPIRou spectrograph at CFHT is providing~\ms~precision spectroscopic observations that extend into the near-IR \citep{artigau2011} where the Zeeman effect is stronger, and will therefore provide excellent observations to improve our techniques to measure unsigned magnetic flux in slowly rotating, quiet Sun-like stars.

\subsection{Prospects improving the $FF'$ and $F^2$ methods} \label{prospects_ff}

In Section~\ref{sec_beyond_ff} and Appendix~\ref{app_ffmethod}, we noted that the $FF'$ and $F^2$ terms of \citet[][]{Aigrain:2012} have intrinsic limitations, largely because they introduce extra limb-darkening-like 
damping to the predicted \drv.  
It may be possible to improve these methods by applying appropriate
``anti-limb-darkening" functions to each, timed to the central
meridian passage a magnetic feature. 
We provide further details in Appendix~\ref{app_improveff}.
There are several difficulties associated with our proposed correction, but we plan to experiment with these ideas in the near future.

\subsection{Future improvements to SDO/HMI-RV pipeline}\label{sec_beyond_code}
The HMI instrument is not stable over timescales longer than a few days, because it was originally designed for helioseismic observations (see Section~\ref{sec_dopplergrams} and Figure~\ref{fig_rv_hat}). To correct for this, we look at RV \emph{variations}, which we obtain by subtracting the velocity of the quiet Sun. In the present analysis, we estimate this quiet-Sun velocity by excluding magnetically active pixels. This is a reasonable approximation since active regions occupy only a few \% of the solar disc (5\% at the peak of the magnetic cycle). 
Ideally, one should replace each active pixel by a non-active pixel from the same spatial location (e.g., from a few days before or after the active region's presence).  This could be achieved by compiling a quiet-Sun ``template" image.
Such a fix is unlikely to account for the rotationally modulated RV residuals we are seeing, but may improve the accuracy of \drv~and ultimately enable us to probe deeper into the physical processes at play in RV variations.

\section{Conclusions}\label{sec_conclusion}

In this paper, we estimate the disc-averaged, rotationally modulated radial-velocity (RV) variations of the Sun as a star over magnetic cycle 24 from spatially resolved images of the Sun taken by the Helioseismic and Magnetic Imager onboard the Solar Dynamics Observatory (SDO/HMI). 
To do so, we apply a model that was previously validated against overlapping HARPS-N solar observations by \citet{milbourne}.
We also estimate the disc-averaged, unsigned (\ie~unpolarised) magnetic flux.  
Our findings and conclusions are summarised here:
\begin{itemize}
    \item[-]
    The SDO/HMI-derived RV dataset presented here has high cadence and timespan that covers nearly an entire magnetic cycle, and high SNR (Figure~\ref{fig_all}). It thus provides a testbed to identify and probe the underlying physical processes that are responsible for rotationally modulated RV variations.
    
    \item[-] Periodograms of the Sun's RV (Figure~\ref{fig_GLS}) show that the majority of the power is shared between the rotation period ($P$) and its first harmonic ($P/2$). Both peaks are significant, and each are in fact broad forests of significant peaks that lie up to 2-3 days away from $P$ and $P/2$. 
    
    \item[-] We fit RV variations with a linear model of the unsigned magnetic flux, and find that it reduces the \rms~of RV variations by 62\%  \ie~a factor of 2.6, from 2.33~\ms~to 0.89~\ms~(Figure~\ref{fig_model_b}, Section~\ref{sec_model_b}). The residuals of the fit display rotationally modulated behaviour, particularly at times of high magnetic activity (Figure~\ref{fig_zoom}). To try to account for these residuals, we fit RV variations with a combination of a linear \bhat~term and an \ff~term from the method of \citet{Aigrain:2012} (Figure~\ref{fig_model_bff}, Section~\ref{sec_model_bff}). This yields only modest \rms~improvements, as the combined model gives a residual \rms~of 0.85~\ms~for the full timeseries. 
We show that the \ff~model does not adequately account for RV variations from magnetic regions because it over-accounts for limb darkening (Section~\ref{sec_others} and Appendix~\ref{app_ffmethod}), and we propose a correction to potentially improve the performance of the \ff~method (Section~\ref{prospects_ff} and Appendix~\ref{app_improveff}).
    
    \item[-] Modelling RV variations with \bhat~and the \ff~method allows us to identify additional physical processes responsible for rotationally modulated RV variations.
    These signals are either missing from the RV model of \citet{milbourne} and \citet{haywood2016} that we use to estimate RV variations from SDO/HMI images, or they are not well traced by \bhat~or \ff, or both.
    Particularly at high magnetic activity levels, the residuals display significant, rotationally modulated variations at the meter-per-second-level.
We discuss physical processes that may contribute to these additional RV variations beyond suppression of convective blueshift and brightness inhomogeneities: horizontal flows (such as Evershed flows, moat flows and active region inflows), flows that are not resolved by SDO/HMI's pixels, and Zeeman broadening (Section~\ref{sec_others}).
    
    \item[-] We inject planet signals to test the performance of the unsigned magnetic flux \bhat~for mitigating rotationally modulated RV variations in surveys of low-mass, long-period planets orbiting Sun-like stars (Section~\ref{sec_planet}). 
    We inject planets with orbital periods of $\simeq$~300~days and RV semi-amplitudes of 0.5, 0.3 and 0.1~\ms.
    The \bhat~model and the combined \bhat~+~\ff~model give very similar results.
    The parameters of the planets with $K$~=~0.3 and 0.5~\ms~are detected accurately to within 1-$\sigma$ of the injected parameters.
    We do not retrieve injected signals with $K$~=~0.1~\ms, because of the presence of an activity-induced signal at 330 days.
  \end{itemize}
We conclude that \bhat~could, in principle, enable us to extract planet signals down to 0.3~\ms, but we will also need to model additional RV variations to reach 0.1~\ms~(Section~\ref{sec_stars}). The significance of planet detections in stellar observations will depend crucially on how precisely we may be able to measure \bhat. Stellar RV observations will also be affected by (super)granulation signals at the \ms~-level \citep{meunier2015}, instrumental systematics, and ground-based observing schedules. 
The most promising avenue to measure the unsigned magnetic flux in slowly rotating, relatively inactive stars is by measuring Zeeman broadening of magnetically sensitive lines in high-precision spectra \citep[\eg][]{Kochukhov20}.

\facilities{SDO/HMI; SORCE; HARPS-N; Mount Wilson Observatory HK Project.}  

\acknowledgments
We are grateful to Colin Folsom and Adriana Valio for insightful discussions on magnetic flux.
This work was performed under contract with the California Institute of Technology (Caltech)/Jet Propulsion Laboratory (JPL) funded by NASA through the Sagan Fellowship Program executed by the NASA Exoplanet Science Institute (R.D.H.).
This work was supported in part by NASA award number NNX16AD42G and the Smithsonian Institution (T.W.M.).
A.M. acknowledges support from the senior Kavli Institute Fellowships.
A.C.C. acknowledges support from STFC consolidated grant number ST/M001296/1.
S.H.S. is grateful for support from NASA Heliophysics LWS grant NNX16AB79G.
H.M.C. acknowledges the financial support from a UK Research and Innovation Future Leaders Fellowship, as well as support from the National Centre for Competence in Research PlanetS supported by the Swiss National Science Foundation.
This work is made possible by a grant from the John Templeton Foundation. The opinions expressed in this publication are those of the authors and do not necessarily reflect the views of the John Templeton Foundation. 
This material is based upon work supported by the National Aeronautics and Space Administration under Grant No. 80NSSC18K0476 issued through the XRP Program.
We gratefully acknowledge the support of the international team 453 by the International Space Science Institute (Bern, Switzerland).
The HMI data used are courtesy of NASA/SDO and the HMI science team. SDO is part of the Living With a Star Program within NASA's Heliophysics Division. 
This research has made use of NASA's Astrophysics Data System and the NASA Exoplanet Archive, which is operated by the California Institute of Technology, under contract with the National Aeronautics and Space Administration under the Exoplanet Exploration Program.

\bibliography{sample63}{}
\bibliographystyle{aasjournal}

\appendix
\restartappendixnumbering

\section{Hystereses between \drv, S-index, and \bhat}\label{app_hysteresis} 
\citet{meunier2019c} showed that for a given level of \ca~emission, RV variations have a comparatively lower amplitude during the descending phase of the magnetic cycle than in its ascending phase. 
This is likely because the average latitude $\langle |\phi |\rangle$ of active regions changes over the course of the magnetic cycle, highest in the ascending phase, and lowest in the descending. 
Signals produced by an active region depend on the region's position on the solar disc; RV and S-index behave differently with line-of-sight angle and follow different limb-darkening laws, because they originate at different heights in the solar atmosphere \citep[further details in][Sect.6.1]{meunier2019c}. Note that \bhat~is a line-of-sight observable, and only subject to foreshortening.  

\begin{figure*}[t]
\centering
\includegraphics[scale=0.5]{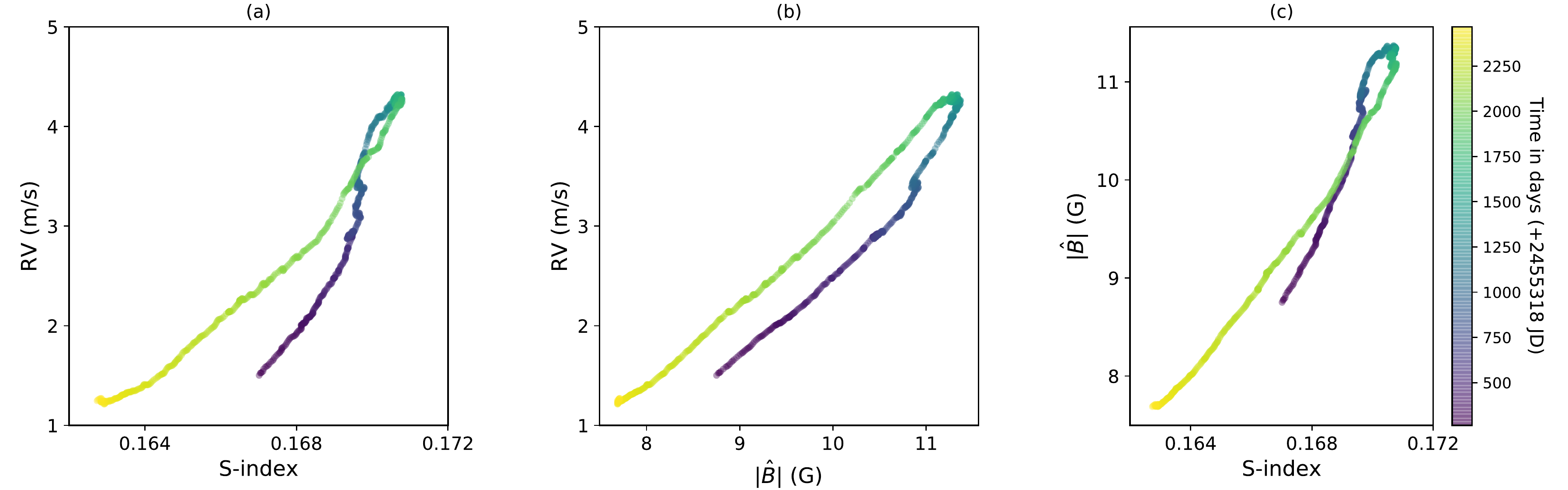}
\caption{Hysteresis between RV variations and: (a) \ca~emission as measured by the S-index, (b) unsigned magnetic flux \bhat~and (c) between \bhat~and the S-index. All timeseries are smoothed over 300-day bins to reveal the long-term hystereses. The points are colour-coded according to time: the ascending phase (increasing activity) is in purple while the descending phase (decreasing activity) is in green and yellow. \label{fig_hysteresis}}
\end{figure*}

To examine these long-term effects, we averaged our timeseries with a 300 day boxcar smooth.  This is long enough to smooth out both rotational modulation, and the growth-decay timescales of large active regions, thus concentrating on purely cyclic variation.  
As shown in Figure~\ref{fig_hysteresis}, we observe a hysteresis between \drv~and S-index (panel (a)), between \drv~and the unsigned magnetic flux \bhat~(panel (b)), and between the S-index and \bhat~ (panel (c)).

The RV--S-index hysteresis looks qualitatively similar to that observed in the previous solar magnetic cycle by \cite[][Fig.7]{meunier2019c}.  In the ascending phase (dark purple line in Figure~\ref{fig_hysteresis}), plages, at higher $\langle |\phi |\rangle$, are (more) limb-brightened, and also increasing in total area, yielding a larger S-index per unit projected plage area than later in the cycle.  
There is also an RV effect, as the difference in RV between non-magnetic pixels and plage pixels peaks at $\mu \sim 0.9$ \citep{palumbo2017}. This azimuthal ring lies entirely within $\phi = \pm 25^o$, the approximate ``active latitude zone".   Thus, as the cycle progresses and active-region average latitude decreases, the per-pixel average RV difference with the quiet Sun {\it decreases}. In the ascending phase, 
this is more than compensated by the increasing filling factor, but once $f_{\rm plage}$ starts declining in the descending phase, \drv~drops steadily.   

The hysteresis between \drv~and \bhat~is in some ways simpler to understand. At the 300 day level of smoothing, the twin cycle maxima of cycle 24 at $t \sim 500$ days (the weaker Northern hemisphere peak) and $t\sim 1600$ d (the stronger Southern hemisphere peak) are both flattened into one slow increase in \bhat.  With an initial $\phi \sim 25^o$, then mostly decreasing throughout the cycle, the average net RV per plage pixel should also be decreasing.  This is counterbalanced, however, by the filling factor, which increases more quickly than RV decreases in both the ascending phase and the peak(s) of the cycle.  Thus, \drv~continues to increase with \bhat~in these phases, more slowly at maximum when \bhat~growth is also reduced, then only finally reversing in the decline phase, when both \bhat~and \drv~plummet.  

Panel (c) of Figure~\ref{fig_hysteresis} shows a hysteresis between S-index and \bhat, which is likely due to the different limb-darkening behaviours, with the S-index getting a boost from limb brightening when in the higher $\langle |\phi |\rangle$ ascending phase. 

Further study is needed to better understand the differences in projection effects between these three observables, to correct for them and therefore obtain tighter correlations between \drv, S-index and \bhat.

\paragraph{Attempt to account for the hysteresis between \drv~and \bhat}
To capture the information in the hysteresis of Figure~\ref{fig_hysteresis}~(b), we fitted the ascending and descending phases of the magnetic cycle separately, using two \bhat~terms and two zero-point offsets. The magnetic cycle has a double-peaked shape, because the active region bands reach maximum activity levels at slightly different times. We identified the peak of the magnetic cycle, \ie~the point separating the ascending and descending phases as the minimum in magnetic flux and active-region coverage between these two peaks, at JD = 2456957.5 (day 1639 of the timeseries shown in  Figures~\ref{fig_all} and~\ref{fig_model_b}). 
We obtain different model parameters for the ascending ($\alpha$ = 11.15~$\pm$~0.01~\ms, $RV_0$ = -7.98~$\pm$~0.01~\ms) and descending phases ($\alpha$ = 8.64~$\pm$~0.01~\ms, $RV_0$ = -6.26~$\pm$~0.01~\ms).
However, the residual \rms~(over the full cycle) is 0.83~\ms~(64\% reduction in RV variations), which is a only a small improvement compared to fitting the full cycle as one (\rms~=~0.89~\ms, 62\% reduction).
This is an improvement, but since it is small and adds more parameters and complexity, we leave this avenue open for future investigations.

\section{Correlations between filling factors and activity indicators}
\label{app_correlations}
\begin{figure*}[]
\centering
\includegraphics[scale=0.65]{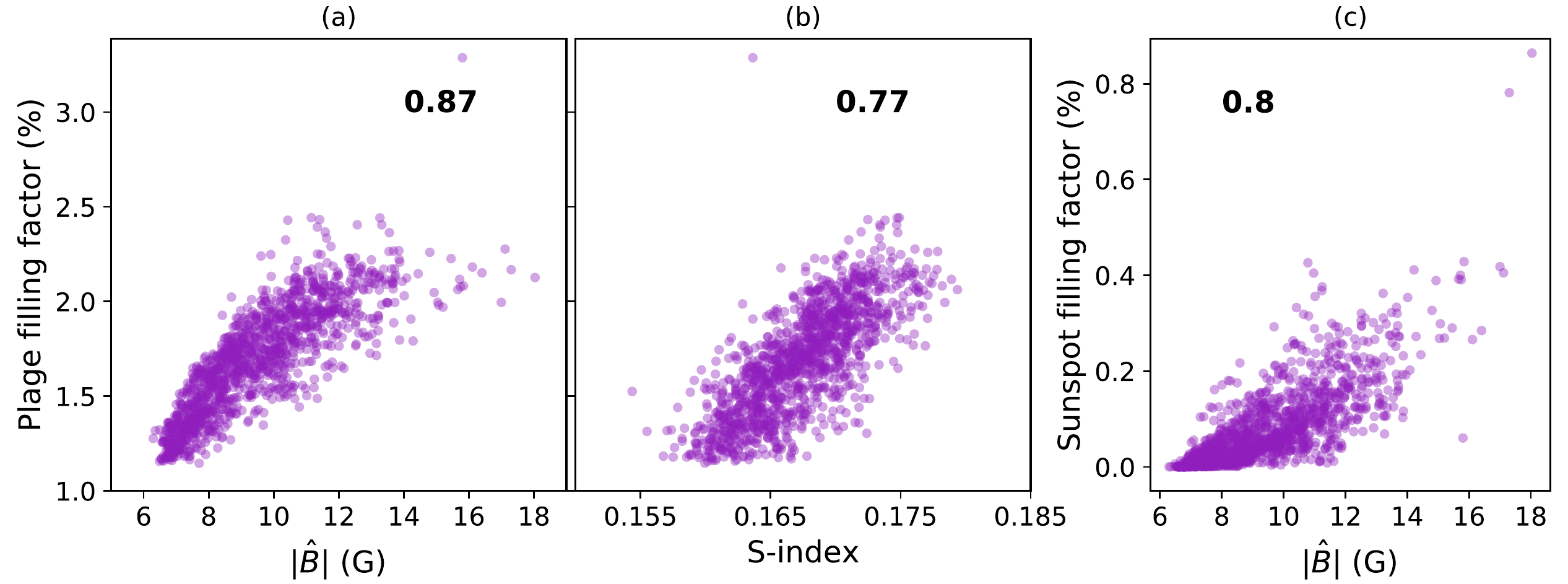}
\caption{Plage filling factor plotted against: (a) \bhat~and (b) S-index. (c) shows the sunspot filling factor as a function of \bhat. Spearman correlation coefficients are given for each pair of correlates. \label{fig_ffactors}}
\end{figure*}

In Figure~\ref{fig_ffactors}, we show the plage filling factor as a function of \bhat~and S-index, and the spot filling factor as a function of \bhat. The magnetic filling factors are estimated according to Eqn.~\ref{eqn_ff}.

\section{Demonstration of the limitations to the \ff~method}
\label{app_ffmethod}
Here, we show why the \ff~method  cannot match RV variations perfectly.
Consider an equatorial
spot. If we assume, for simplicity, linear limb darkening, and solar inclination $i=90^o$, the flux from the spot of area $A$ can be written as:
\begin{equation}
F = A \cos \theta (1- \epsilon_{\rm s} +\epsilon_{\rm s} \cos \theta) ,
\end{equation}
where $\theta$ is the angle of the surface normal to the line of sight (at
disk center), and $\epsilon_{\rm s}$ is the linear limb darkening
coefficient for the spot.  
The derivative with respect to $\theta$ is:
\begin{equation}
F' = - A \sin \theta (1- \epsilon_{\rm s} + 2\epsilon_{\rm s} \cos \theta) 
\end{equation}
To match
the RV change due to the rotation of a spot, this simple model \citep[based on][]{saar1997} yields:
\begin{equation}
\Delta v_{\rm spot} = - A \sin \theta \cos \theta (1- \epsilon_{\rm s}
+\epsilon_{\rm s} \cos \theta)  .
\end{equation}
The $\sin \theta$ term captures the RV deficit, and is weighted
by the projected area of the spot ($\cos \theta$) and its limb
darkening, $(1- \epsilon_{\rm s}
+\epsilon_{\rm s} \cos \theta)$.  

In comparison, the $FF'$ method delivers:
\begin{equation}
FF'/A = -A \sin \theta \cos \theta (1- \epsilon_{\rm s} +\epsilon_{\rm
s} \cos \theta) (1- \epsilon_{\rm s} + 2\epsilon_{\rm s} \cos \theta), 
\end{equation}
or, 
\begin{equation}
F F'/A = \Delta v_{\rm spot} (1- \epsilon_{\rm s} + 2\epsilon_{\rm s} \cos \theta).
\end{equation}
Thus, the $FF'$ method captures $\Delta v_{\rm spot}$ 
but adds an additional limb-darkening-like term.  
This leads
to systematic effects, underestimating $\Delta v_{\rm spot}$
 progressively more and more as the spot moves away from disk center.

Although we do not use the $F^2$ term proposed by \citet[][]{Aigrain:2012} to correct for the convective suppression arising primarily in plage, we note that it is similarly flawed.  Following a similar analysis:
 \begin{equation}
 \Delta v_{\rm plage} =  A \cos^2 \theta (1- \epsilon_{\rm p} +\epsilon_{\rm p} \cos \theta)  ,
 \end{equation}
 where $\epsilon_{\rm p}$ is the linear limb darkening coefficient
 for plage.  But $F^2$ yields: 
 \begin{equation}
 F^2/A = A \cos^2 \theta (1- \epsilon_{\rm p} +\epsilon_{\rm p} \cos \theta)^2  ,
 \end{equation}
 or, 
 \begin{equation}
 F^2/A = \Delta v_{\rm plage} (1- \epsilon_{\rm p} +\epsilon_{\rm p} \cos \theta) ,
 \end{equation}
 which contains an extra limb-darkening term.   

In the case of a spot, its limb darkening $\epsilon_{\rm s}$ is unlikely to differ significantly from the 
quiet Sun value $\epsilon_{\rm q}$.  This is because limb darkening is wavelength dependent, and the strength-weighted average 
for HARPS-N RV lines is perhaps $\langle \lambda \rangle \sim 500$ nm, not much different from the HMI line ($\lambda$ = 617.3 nm).  
\ff~captures the RV perturbation due to a spot crossing the disc effectively, but then applies a second,additional, stronger limb darkening. 

We thus warn that, while the \ff~term is partially successful, it (and the related $F^2$ term)  are also flawed, since they
intrinsically add extra, unwanted limb-darkening-like corrections.
Therefore, one cannot expect the \ff~term  (or the $F^2$ term) to perfectly account for the RV variations of magnetic features.

\section{On improving the \ff~and $F^2$ terms}
\label{app_improveff}
For passage of a spot (detected, \eg~in photometry), the corrected
$FF'$ becomes:
\begin{equation}
FF'_{\rm corrected,spot} = FF'/(1 - \epsilon_{\rm s} + 2 \epsilon_{\rm s} \cos \theta). 
\end{equation}
A similar formula can similarly be applied to plage (detected, \eg~in \ca), by using the limb darkening coefficient $\epsilon_{\rm p}$.
%
%
We note that the equivalent correction to $F^2$ for convective suppression, which occurs predominantly in plage (rather than spots) is: 
\begin{equation}
F^2_{\rm corrected,plage} = F^2/(1 - \epsilon_{\rm p} +  \epsilon_{\rm p} \cos \theta) .
\end{equation}
In order to apply the above corrections, one has to know when individual active regions cross the solar/stellar disc.
To some extent, one can track the meridian passage of active regions on solar/stellar surfaces via monitoring of disc-averaged photometric and \ca~emission.
However, an active region often contains both spots {\em and} plage, resulting in a mixed, degenerate photometric signal. Also, it would be difficult to apply this correction in terms of timing when multiple active regions are present.
Additionally, we note that $\epsilon_{\rm s}$ and  $\epsilon_{\rm
p}$ remain poorly known due to lack of realistic models. Better values may be derivable from 3D-MHD simulations \citep[e.g., ][]{cegla2018}. It is also difficult to 
choose an appropriate value of $\langle \epsilon \rangle$ for spectra that span hundreds of nm. 
To summarise, applying a correction to the \ff~term could potentially improve fits to RV variations incurred by spots and plage, but there will likely remain residual RV variations.

\end{document}